\documentclass[preprint,3p]{elsarticle}

\usepackage{color}
\usepackage{epsf}
\usepackage{graphicx}
\usepackage{parskip}            % Abs‚Ä∞tze durch Zeilenabstand trennen
\usepackage{psfrag} 
\usepackage{xspace}
\usepackage{paralist}
\usepackage{url}
\usepackage{hyperref}
\usepackage{tikz}				% SchÀÜne Bildchen!

\usepackage{geometry}
\graphicspath{{fig/},{pdf/},{eps/}}
\usepackage[utf8]{inputenc}
\usepackage{listings}
\usepackage{float}

\floatstyle{boxed}
\newfloat{sidebar}{thbp}{sb}
\floatname{sidebar}{Sidebar}

\begin{document}

%%%%%%%
%%%%%%%
\begin{frontmatter}

%\listoftodos

%\floatstyle{boxed}
%\newfloat{sidebar}{thbp}{sb}
%\floatname{sidebar}{Sidebar}

%\begin{titlepage}

\title{Analysing Text in Software Projects}

\author[label1]{Stefan Wagner\corref{cor1}}
\ead{stefan.wagner@informatik.uni-stuttgart.de}
\author[label2]{Daniel M\'{e}ndez Fern\'{a}ndez}

\address[label1]{Software Engineering Group, Institute of Software Technology, University of Stuttgart, Universit\"atsstr.~38, 70569 Stuttgart, Germany}

\address[label2]{Software \& Systems Engineering, Institut f\"ur Informatik, Technische Universit\"at M\"unchen, Boltzmannstr.~3, 85748 Garching, Germany}

\cortext[cor1]{Corresponding author}

%\author{Stefan Wagner (University of Stuttgart, Germany) \and Daniel M\'{e}ndez Fern\'{a}ndez (Technische Universit\"at M\"unchen, Germany)}

%Fakult\"at f\"ur Informatik &  Fakult\"at f\"ur Informatik, Elektrotechnik & Fakult\"at f\"ur Informatik    \\ 
% Technische Universit\"at M\"unchen	&und Informationstechnik & Technische Universit\"at M\"unchen	    \\
%&Universit\"at Stuttgart    &   \\

%\maketitle

%\end{titlepage}

\begin{abstract}
Most of the data produced in software projects is of textual nature: source code, specifications, or documentations. The advances in quantitative analysis methods drove a lot of data analytics in software engineering. This has overshadowed to some degree the importance of texts and their qualitative analysis. Such analysis has, however, merits for researchers and practitioners as well. 

In this chapter, we describe the basics of analysing text in software projects. We first describe how to manually analyse and code textual data. Next, we give an overview of mixed methods to automatic text analysis including N-Grams and clone detection as well as more sophisticated natural language processing identifying syntax and contexts of words. Those methods and tools are of critical importance to aid in the challenges in today's huge amounts of textual data.

We illustrate the introduced methods via a running example and conclude by presenting two industrial studies.


\end{abstract}

\end{frontmatter}

%\newpage

%\tableofcontents

%\newpage

\section{Introduction}% (SW)(300 words, 1 fig)}

Most of the data we produce in software projects is of textual nature. It starts with requirements specifications over designs and documentation to customer surveys. 
Textual data, however, is notoriously difficult to analyse. We have a multitude of techniques and tools which are well known to software developers
to process and analyse quantitative data. We handle numbers of bugs or lines of code very often. Yet, what should we do with all the textual data? So far,
the potential of analysing this data is not used.

Hence, software analytics for practitioners does already (and will even more) involve how to make use of all this textual data. Fortunately, we have seen
an interesting development in the research area of analysing texts. This is often promoted under the umbrella \emph{text mining} or \emph{text analytics}.
Both mean roughly the same thing: systematically retrieve and analyse textual data to gain additional insights, in our case additional insights into the
software development project.

In this chapter, we will first discuss and categorise where and what kind of textual data we usually encounter in software projects. Based on this categorization,
we will discuss the sources of this textual data and how to retrieve them. Next, we will introduce manual coding and analysis as a very flexible but elaborate
means to structure and understand different texts. As a means to handle large amounts of text and reduce the manual effort, we will discuss a sample of the nowadays available automatic analyses of texts such as N-Grams or clone detection.

Finally, we will employ a running example of textual data which we will analyse with each presented analysis approach. We use the publicly available 
specifications of the  HTTP and IMAP protocols as typical representatives for textual requirements specifications. In case of automatic analyses, we will
provide references to the used tools so that the examples can easily be reproduced.

\section{Textual Software Project Data and Retrieval}
\label{sec:categories} 

Textual data appears at many points in a software development project. The main result of a software project -- the code -- is also textual although we are more concerned
with natural language text in this chapter. Yet, source code usually contains also a considerable share of natural language text: the code comments. Apart from that, we have
requirements specifications, architecture documentation or change requests with textual data. There is no generally agreed on classification of the artifacts generated in a
software development project and, hence, also not about textual data contained in these artifacts. As a foundation for the remainder of this chapter, we will therefore first discuss different sources for textual data and then classify them to describe how to retrieve the data for further analysis.

\subsection{Textual Data}

\paragraph{Sources for Textual Data}
Although we often consider software developers to be people dealing mostly with formal languages, i.e.\ programming languages and alike, there are many sources for textual
data in a software project. Most of this data comes from the developers themselves. A substantial part of any high-quality source code consists of
informal or semi-formal text code comments. These already contain smaller inline comments as well as longer descriptions of interfaces or classes (as e.g.\ with JavaDoc for Java).
Apart from that, depending on the development process followed, there are many other documents written by the developers or other people in related roles such as testers, software
architects or requirements engineers. This then includes requirements specifications, design and architecture documents, test cases, test results and review results. Yet, we need to think also beyond the ``classical'' artifacts in software development. Team members nowadays communicate via e-mails and chats, change requests and commit messages which all are available electronically as text sources. Finally, in a software project, especially in the context of product development, customer surveys and product reviews are a common means to better understand the requirements for and satisfaction with a software product. These artifacts can contain also open questions which result in textual data to be analysed.

\begin{figure}[htb]
\centering
  \includegraphics[width=1\textwidth]{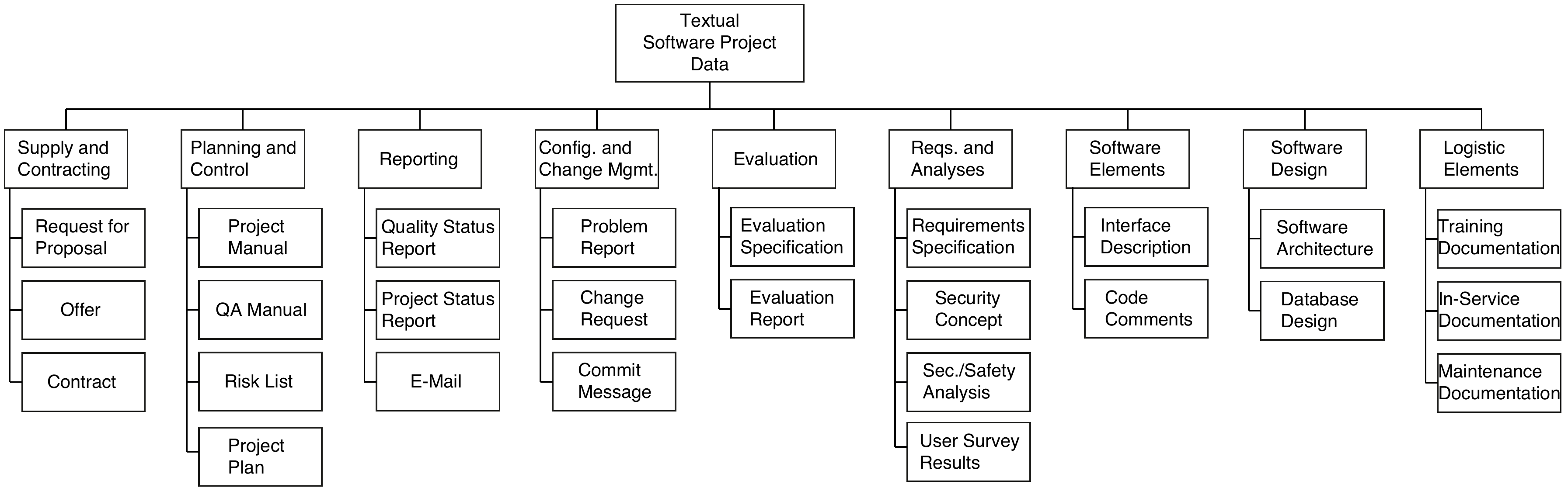}\\
  \caption{A classification of textual software project data (based on \url{http://www.v-model-xt.de/})}\label{fig:classification}
\end{figure}

\paragraph{A Classification of Textual Software Project Data}
%We do not have a commonly agreed on classification of artefacts created in software engineering. This is a problem because of the many different 
%development processes in use and the corresponding differences in the created artefacts. 
We found the general structure of the German standard software process model \emph{V-Modell XT}\footnote{\url{http://www.v-model-xt.de/}} to be useful to classify artefacts related to software engineering as it is rather general and applies to many application domains. It contains several detailed artefacts that we do not consider in our classification of textual software project data, but we follow the general structure. The classification is shown in Figure~\ref{fig:classification}.

The \emph{V-Modell XT} structures the textual artefacts mainly along different process areas or disciplines in the development process. This starts with documents for \emph{contracting} such as
requests for proposals or offers. One could, for example, be interested in analysing public requests for proposals to investigate technological trends. Next, there are several
textual documents in project \emph{planning and control}. An example that could be interesting for text analytics are risk lists from projects where we could extract typical risk topics to define mitigation points. We classified the above mentioned e-mails under \emph{reporting} which could give relevant insights into the communication between developers or between developers and
customers. A now commonly analysed area in software engineering research is \emph{configuration and change management}. Problem reports, change requests and commit
messages all contain valuable textual information about a project's progress. With \emph{evaluation}, we mean artifacts related to evaluating other artifacts in the project, for example
test case specifications. We could analyse, for instance, the terminology used in the test cases and compare it to the terminology used in the code. The next three categories correspond to the constructive phases of the project: \emph{requirements and analysis}, \emph{software design} and \emph{software elements}. All contain valuable information about the product. In the latter, we see also the code and with it the code comments in which we can check the used language by text analytics. Finally, the \emph{logistic elements} contain any further documentation which could also be analysed for the discussed topics and for relationships between documents.

\paragraph{Running Example}
As a running example for the remainder of this chapter, we chose the specifications of internet protocols. They are openly available examples of software requirements specifications. The IETF publishes all open protocols as so-called \emph{Requests for Comments} (RFC) and, hence, these documents are called RFC XXXX where ``XXXX'' is the number
of the specification. We selected standards around two well-known internet protocols: HTTP and IMAP. We hope that by choosing these, we avoid lengthy introductions and
potential misunderstandings about the domain. HTTP (Hypertext Transfer Protocol) is the application-level internet protocol for most network applications today. IMAP (Internet
Message Access Protocol) allows access to mailboxes on mail servers. For methods that require a larger corpus, we add additional RFCs to the
corpus that are related to HTTP or IMAP.

The following is a part of RFC 2616 of the specification of HTTP 1.1. It describes valid comments in the HTTP header. It shows that our examples contain text
that is similar to text in other requirements specifications:

\begin{quotation}
Comments can be included in some HTTP header fields by surrounding
   the comment text with parentheses. Comments are only allowed in
   fields containing "comment" as part of their field value definition.
   In all other fields, parentheses are considered part of the field
   value.
\end{quotation}

Yet, also other kinds of text need to be analysed. In our industrial studies in Section~\ref{sec:IndustrialStudies}, we see that free-text answers in surveys are usually not well-formed, complete sentences. Also code comments are often not complete sentences. Although we will not discuss this in detail in the following, most techniques are able to cope with this.

\subsection{Text  Retrieval}% (DMF) (900 words, 1 table, 1 fig)}
\label{sec:TextCollection}

\begin{figure}[htb ]
\centering
  \includegraphics[width=.7\textwidth]{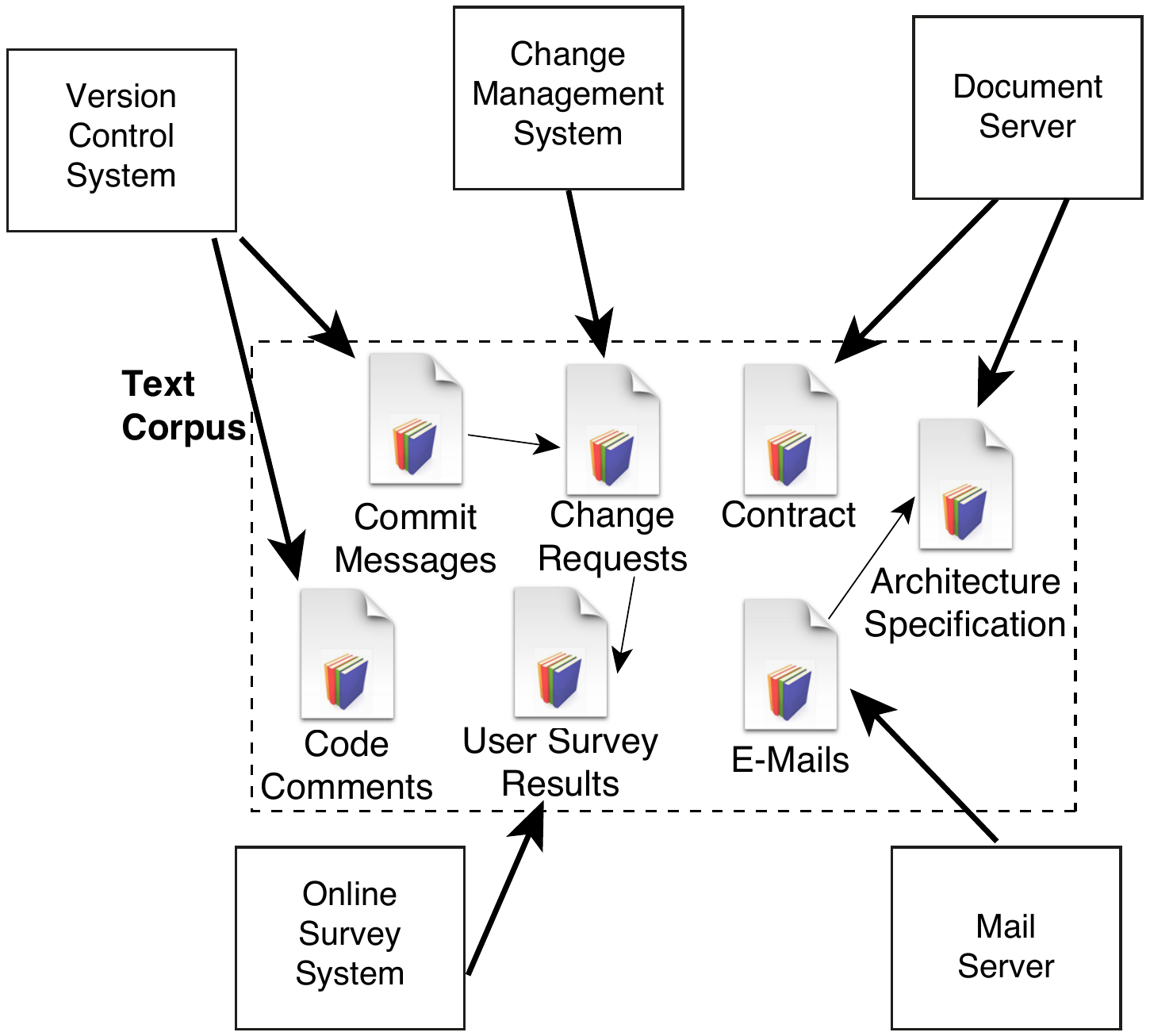}\\
  \caption{Text Collection from Different Sources}\label{fig:text-collection}
\end{figure}

With the various sources and classes of different texts in software engineering projects, the first step is to collect or retrieve these texts 
from their sources. A complete text corpus, consisting of different texts, will often come from different sources and we want to keep connections between the texts as stored in the sources. For example, we often have pointers from the commit messages in the version control system to specific change requests in the change management system. These links are useful for further analysis and should be retrieved. Figure~\ref{fig:text-collection} gives an overview of such retained links between texts in the text corpus. In the following, we will go through the categories of textual project data from Section~\ref{sec:categories} and discuss the sources and possible
links between texts.

The texts in the supply and contracting category are often not stored with most of the other project documentation. They are usually held either in the form of
a formatted document on a file server or part of a larger enterprise resource planning (ERP) system. Hence, we either need to access the file server and extract the
plain text from the formatted document or access the ERP system which usually has some kind of API to retrieve data. We should aim to keep links to the actual
project in terms of project IDs or something similar.

Similarly, planning and control texts are also often kept as formatted documents on file servers or in ERP systems. Then, they need to be treated in the same way as supply
and contracting texts. They have a better chance, however, to be also kept in a version control system. For example, the above mentioned risk lists can be maintained also as plain text and, therefore, easily handled in Subversion\footnote{\url{http://subversion.apache.org/}} or Git\footnote{\url{http://www.git-scm.com/}}. Then, we can use the APIs of the corresponding version control system for retrieval. It also allows us to retrieve different versions and, hence, 
the history of documents which can be necessary for temporal analyses.

Reporting can be done in formatted documents, ERP systems as well as version control systems. Quality status reports can also be automatically generated by quality control systems integrated with the continuous integration system. For example, SonarQube\footnote{\url{http://www.sonarqube.org}}
or ConQAT\footnote{\url{http://www.conqat.org/}} can be used to auto-generate quality reports based on data from various sources. These reports can be directly
used if they can be retrieved. Usually they provide the possibility to put the report as a file on a server, into a version control system or into a database. They often
provide links to the artifacts analysed which should be retained if possible. E-mail
can be most easily retrieved if a central mail server is used for the project. While most textual data we retrieve is sensitive, with e-mail we need to take the most
care not to violate privacy rules established in the organisational context. It is advisable to anonymise the data already at this stage and only keep, for example, links to specifications or code explicitly referenced in the e-mails.

Texts from configuration and change management are programatically easy to obtain as they are nowadays stored in databases in most companies. We have the above mentioned version control systems, ticketing systems, e.g.~\emph{OSTicket}\footnote{\url{http://www.osticket.com/}}, and
change or issue management systems such as \emph{Bugzilla}\footnote{\url{http://www.bugzilla.org/}} or \emph{Atlassian Jira}\footnote{\url{https://www.atlassian.com/software/jira}}.
They all provide APIs to retrieve the contained text. Depending on the system and possible naming conventions, we can retrieve, besides the texts, links
between commit messages and change requests or between problem reports and change requests.

The further categories are supposed to be held mostly in version control systems. Exceptions can be evaluation (i.e.\ test or review) reports that sometimes
are stored separately on file servers. Also user survey results can be easiest to retrieve by going directly to the online survey servers used. They often provide
some kind of API or export functionalities. Otherwise, we need to write a web crawler or retrieve it manually. To be able to retrieve code comments, we also need to write an extraction tool that is able to distinguish between code and comments. It can also be useful to distinguish different types of comments as far as this
is possible in the retrieval directly. This could also be a further step in the analysis. 

%Maybe depending on the classification in 2. describing how we can retrieve the data automatically. Maybe something about rules in texts, document templates, commit message rules. Also something about designing questionnaires?

%\subsubsection*{Sampling: Why, When, and How?}

%In case of manual analyses, describe the role of sampling.

\section{Manual Coding}% (DMF)}
\label{sec:manual-coding}

Once we collect textual data for its analysis and interpretation, it needs to be structured and classified. This classification is often referred to as \emph{coding} where we identify patterns in texts, having an explanatory or a exploratory purpose~\cite{BM11} and serving as a basis for further analysis, interpretation and validation. Coding can be done in two ways: manually or automated. In this section, we introduce coding as a manual process. A detailed example for applying the manual coding process is provided in Section~\ref{sec:NaPiRE}.

Although manual coding is often associated with interview research, the data we code is not limited to transcripts as we can structure any kind of textual data given in documents, Wikis or source code (see also Section~\ref{sec:TextCollection}). This kind of structuring is used in social science research and also gains attraction in software engineering research. An approach commonly used in these research areas is Grounded Theory. We briefly describe Grounded Theory in Sidebar~\ref{sb:grounded}, but its theoretical background is not necessary for many practical text analysis contexts.

%Furthermore, although manual coding has its origin in Grounded Theory, we will concentrate for the remainder of this section on the practical side of the actual coding process rather than discussing Grounded Theory in general. 

\begin{sidebar}
\caption{Grounded Theory in a Nutshell}\label{sb:grounded}
Manual coding, as discussed in this chapter, has its origins in Grounded Theory (GT). Because GT is the most cited approach for qualitative data analysis~\cite{BM11} which comes at the same time with a plethora of different interpretations, we briefly clarify its meaning in a nutshell. GT describes a qualitative research approach to inductively build a ``theory'', i.e.\ it aims at generating testable knowledge from data rather than testing existing knowledge~\cite{BM11}. To this end, we thus make use of various empirical methods to generate data, and we structure and classify the information to infer a theory. A theory, in its essence, ``provides explanations and understanding in terms of basic concepts and underlying mechanisms''~\cite{HDT07,WRH+12}. In empirical software engineering, we mostly rely on the notion of a \emph{social theory} \cite{Popper02} and refer to a set of falsifiable and testable statements/hypotheses. As most qualitative research methods, Grounded Theory has its origins in social sciences and it was first introduced in 1967 by Glaser and Strauss~\cite{GS67}.  A detailed introduction into the background of Grounded Theory and the delineation with similar concepts arising along the evolution of Grounded Theory is given by Birks and Miller in~\cite{BM11}. For the remainder of the chapter where introduce a manual coding process, we rely on the terms and concepts as introduced in context of Grounded Theory.
\end{sidebar}

%When applying Grounded Theory, we thus make use of various empirical methods to generate data and we try to structure and classify the information to infer a theory. This theory has further an inductive nature with no apodictic propositions. That is, we try to recognise patterns in our data to form hypotheses, for example based on interview transcripts, which we can subsequently test while its inductive nature, by definition, dictates that the main purpose is to develop the theory rather than testing it, i.e.\ rejecting or confirming/supporting it. As a matter of fact, the analysis of the robustness of a theory is left to subsequent, ideally independent studies.

\subsection{Coding Process}% (500 words, 1 fig)}

Manual coding is a creative process that depends on the experiences, views and interpretations of those who analyse the data to build a hierarchy of codes. During this coding process, we conceptualise textual data via pattern building. We abstract from textual data, e.g.\ interview transcripts or commit comments stated in natural language, and we build a model that abstracts from the assertions in form of concepts and relations. During this coding process, we interpret the data manually. Hence, it is a creative process which assigns a meaning to statements and events. One could also say that we try to create a big picture out of single dots. %\todo{Cite Fred Niedermann (Talk)}

There are various articles and textbooks proposing coding processes and the particularities of related data retrieval methods such as why and how to build trust between interviewers and interviewees (see e.g.\ Birks and Mills~\cite{BM11}). The least common denominator of the approaches lies in the three basic steps of the coding process itself followed by a validation step:
\begin{enumerate}
\item \emph{Open coding} aims at analysing the data by adding codes (representing key characteristics) to small coherent units in the textual data, and categorising the developed concepts in a hierarchy of categories as an abstraction of a set of codes -- all repeatedly performed until reaching a ``state of saturation''.
\item \emph{Axial coding} aims at defining relationships between the concepts, e.g.\ ``causal conditions'' or ``consequences''.
\item \emph{Selective coding} aims at inferring a central core category.
\item \emph{Validation}, finally, aims at confirming the developed model with the authors of the original textual data.
\end{enumerate}

Open coding brings the initial structure into unstructured text by abstracting from potentially large amounts of textual data and assigning codes to single text units. The result of open coding can range from sets of codes to hierarchies of codes. An example is to code the answers given by quality engineers in interviews at one company to build a taxonomy of defects they encounter in requirements specifications. During open coding, we then classify single text units as codes. This can result, for example, in a taxonomy of defect types, such as natural language defects which can be further refined, e.g. to sentences in passive voice. During axial coding, we then can assign dependencies between the codes in the taxonomies. For example, the quality engineers could have experienced that sentences in passive voice have frequently lead to misunderstandings and later on to change requests.The axial coding process then would lead to a cause-effect chain that shows potential implications of initially defined defects. The final selective coding then brings the results of open and axial coding together to build one holistic model of requirements defects and their potential impacts. 

\begin{figure}[!hbt]
\centering
  \includegraphics[width=0.6\textwidth]{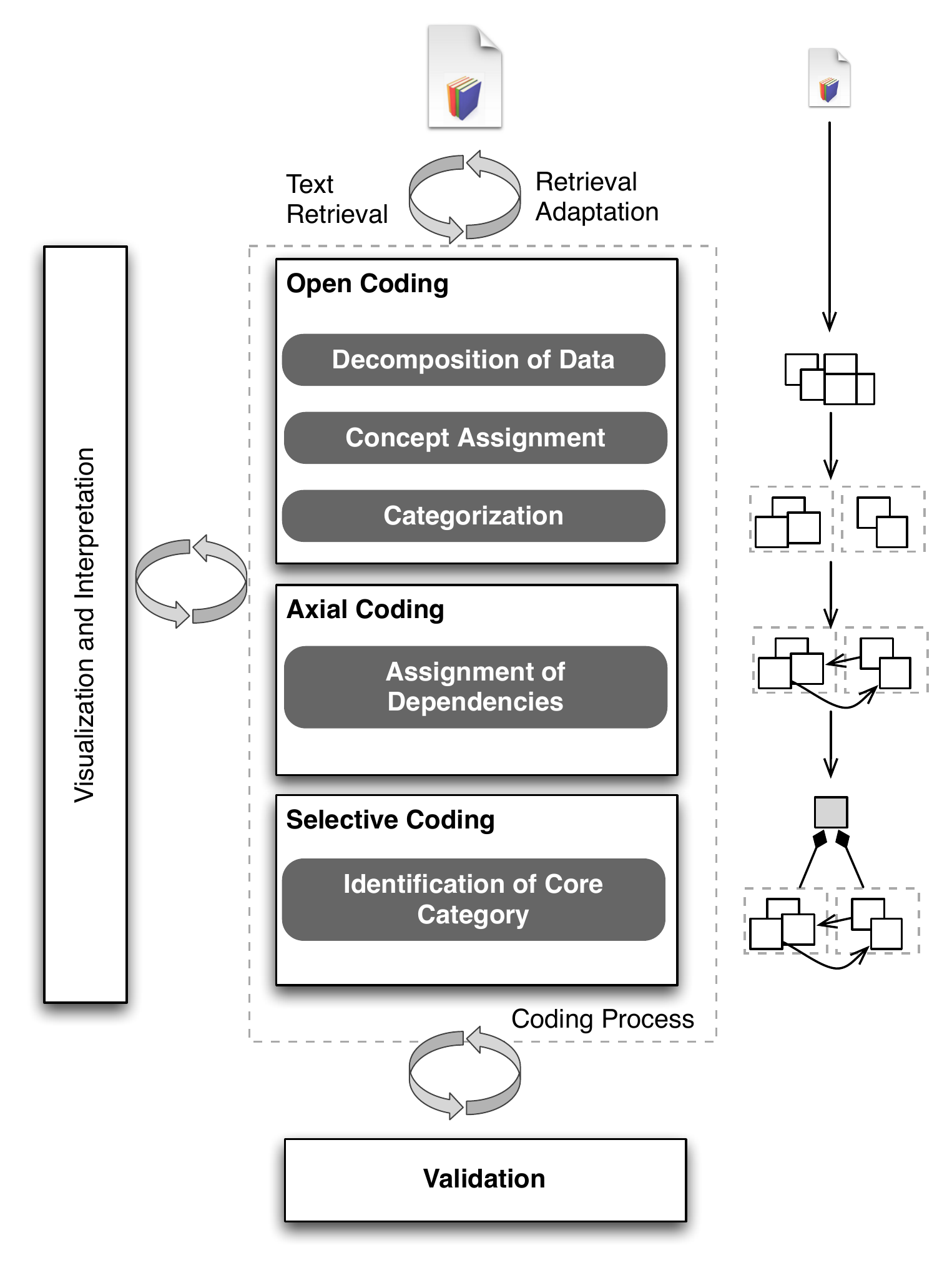}\\
  \caption{Coding process}\label{fig.GTProcess}
\end{figure}

We subsequently form a process that we applied in our studies and which worked well for us. Figure~\ref{fig.GTProcess} depicts the basic coding process and further steps usually (or ideally) performed in conjunction with the coding process. 

The idea of (manual) coding -- as it is postulated in Grounded Theory -- is to build a model based on textual data, i.e.\ ``grounded'' on textual data. As the primary goal is to gather information out of text, we need to follow a flexible process during the actual text retrieval and the coding process as well. For example, in case of conducting interviews, we perform an initial coding of the first transcripts. In case we find interesting phenomena for which we would like to have a better understanding of the causes, we might want to change the questions for subsequent interviews; an example is that an interviewee states that a low quality of requirements specifications has also to do with a low motivation in a team leading to new questions on what the root causes for a low motivation are. We thereby follow a concurrent data generation and collection along with an emerging model which is also steered according to research or business objectives. 

\begin{figure}[htb ]
\centering
\includegraphics[width=.8\textwidth]{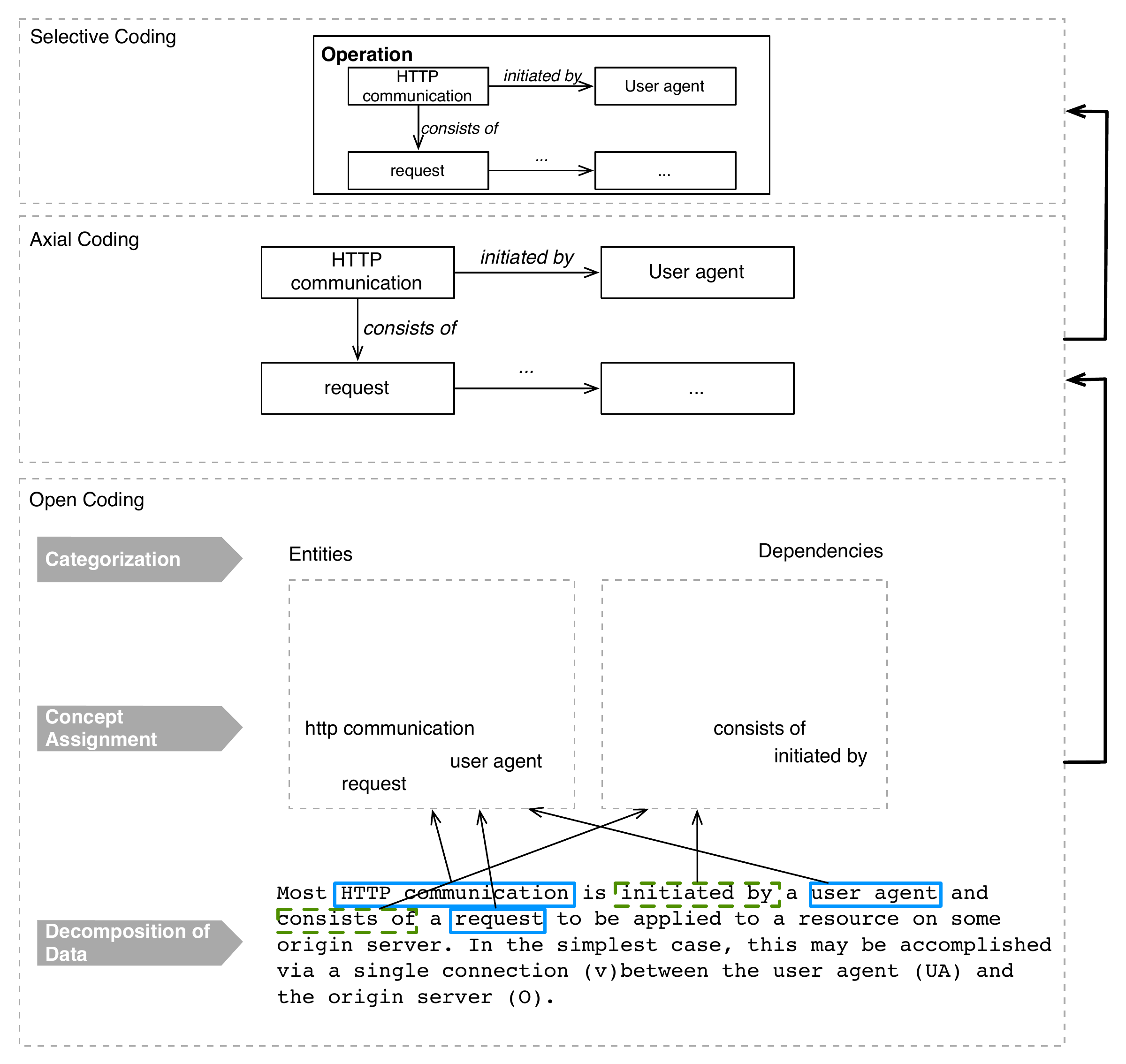}\\
\caption{Manual coding of our running example}\label{fig:manualcoding-example}
\end{figure}

Figure~\ref{fig:manualcoding-example} shows the coding steps for our running example. During the open coding step (lower part of the figure), we continuously decompose data until we find small units to which we can assign codes (``concept assignment''). This open coding step alone shows that the overall process cannot be performed sequentially. During the open coding step, we found it useful
\begin{itemize}
\item to initially browse the textual data (or samples) before coding it to get an initial idea of its content, meaning and, finally, of potential codes we could apply,
\item to continuously compare the codes during coding with each other and especially with potentially incoming new textual data, and 
\item to note the rationale for each code down to keep the coding process reproducible (of special importance if relying on independent re-coding by another analyst).
\end{itemize}

Having a set of codes, we allocate them to a category as a means of abstraction. In our running example, we allocate the single codes to the categories ``entities" and ``dependencies". During axial coding, we then assign directed associations between the codes. Finally, the last step in the coding process is supposed to be the identification of the core category, which often can be also predefined by the overall objective; in our case, it is ``Operation".
%when analysing phenomena in a software development process, we might have various codes for phenomena in a change management process that brings us to simply think in the form of processes so that we might choose ``change management" to be the category. During axial coding, we then assign directed dependencies between the codes. Finally, the last step in the coding process is supposed to be the identification of the core category, which often can be also predefined by the overall objective followed by the text analysis (e.g.\ ``requirements specification defects").

The overall coding process is performed until we reach a theoretical saturation, i.e.\ the point where no new codes (or categories) are identified and the results are convincing to all participating analysts~\cite{BM11}.

\subsection{Challenges}% (100 words)}

The introduced coding process is subject to various challenges, of which we identify the following three to be the most frequent ones.

\paragraph{Coding as a Creative Process}
Coding is always a creative process. When analysing textual data, we decompose it into small coherent units for which we assign codes. In this step, we find appropriate codes that reflect the intended meaning of the data while finding the appropriate level of detail we follow for the codes. This alone shows the subjectivity inherent to coding that demands for a validation of the results. Yet, we apply coding with an exploratory or explanatory purpose rather than with a confirmatory one. This means that the validation of the resulting model is usually left to subsequent investigations. This, however, does not justify a creationist view on the model we define. A means to increase the robustness of the model is to apply analyst triangulation where coding is performed by a group of individuals or where the coding results (or a sample) of one coder are independently reproduced by other coders as a means of internal validation.  This increases the probability that the codes reflect the actual meaning of textual units. We still need, if possible, to validate the resulting model with the authors of the textual data or the interviewees represented by the transcripts.

\paragraph{Coding Alone or Coding in Teams}

This challenge considers the validity of the codes themselves. As stated, coding (and the interpretation of codes) is a subjective process that depends on the experiences, expectations and beliefs of the coder who interprets the textual data. To a certain extent, the results of the coding process can be validated (see also the next paragraph). Given that this is not always the case, however, we recommend to apply, again, analyst triangulation as a means to minimise the degree of subjectivism.

\paragraph{Validating the Results}
We can distinguish between an internal validation, where we form, for example, teams of coders to minimise the threat to the internal validity (the above mentioned analyst triangulation), and external validation. The latter aims at validating the resulting theory with further interview participants or people otherwise responsible for the textual data we interpret. This, however, is often not possible; for example, in case of coding survey results from an anonymous survey. In such cases, the only mitigation we can opt for is to give much attention to the internal validation where we try to increase the reliability of the theory during its construction, e.g. by applying analyst triangulation.

%\todo{What else? In NaPiRE, for example, we only took those results for the new theory that had a minimal occurrence.}
%
%\paragraph{Theoretical Saturation.}
%
%\todo{Saturation is reached when the deadline is coming... ;-)}
%-- Saturation is not predictable... 
%
%Make use of validation 

\section{Automated Analysis}% (SW)}

As any manual coding and analysis of textual data is difficult, largely subjective and very elaborate, automation can have
a huge positive effect. Especially in the recent years, automated natural language processing has made progress we can exploit for analysing software data. 
We cannot replace reading a complete introductory book on natural language processing with this chapter. Yet, we will concentrate on a selected set of promising
techniques and corresponding tools that can give us insight into software engineering data complementing
manual coding.

\subsection{Topic Modelling}% (350 words, 1 fig)}
\label{sec:topic-modelling}

\begin{figure}[hbt]
\centering
  \includegraphics[width=0.9\textwidth]{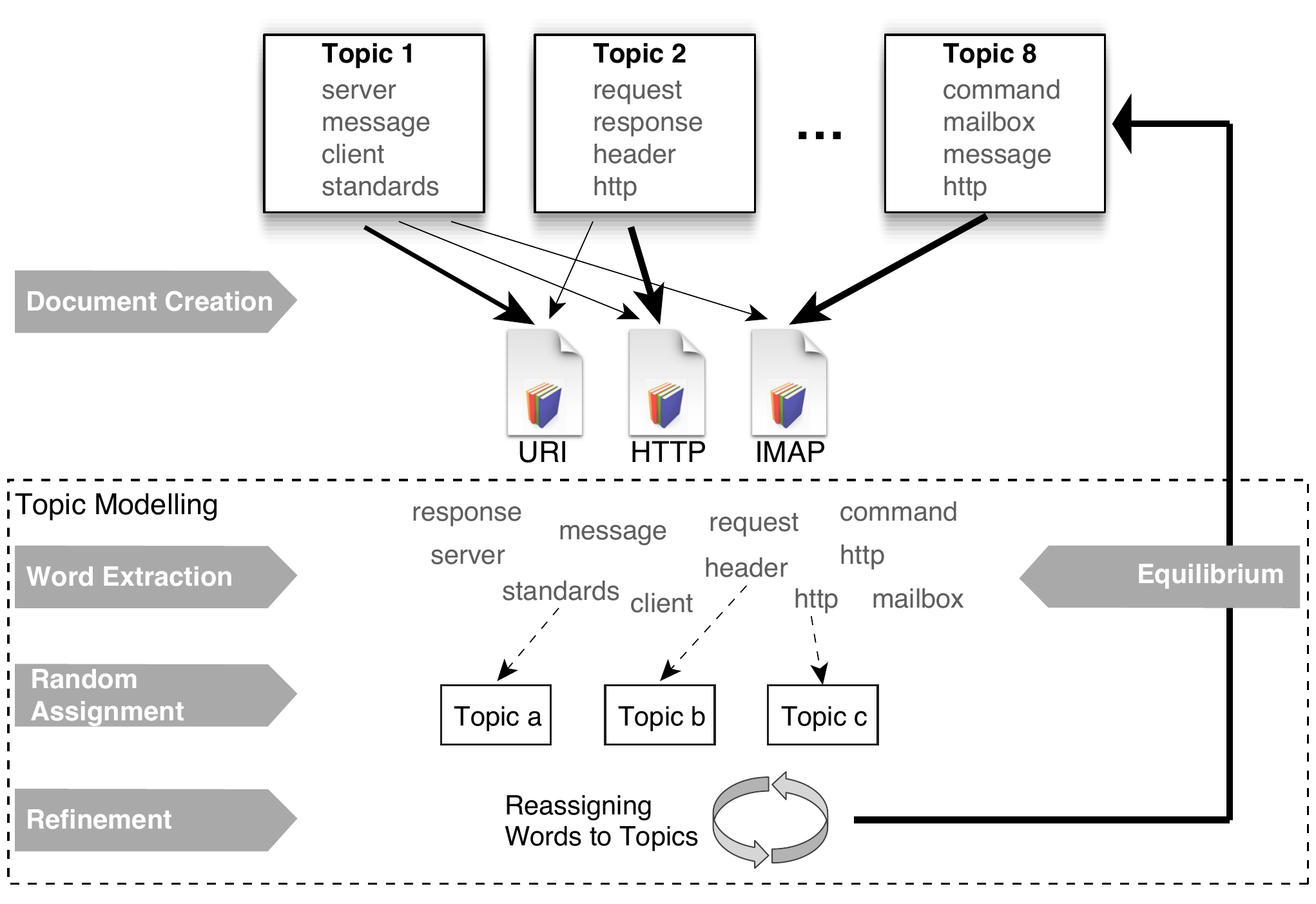}\\
  \caption{Schematic overview of document creation based on topics and topic modelling}\label{fig:topic-modelling}
\end{figure}

We often want to get a fast overview of what different texts are about, for example to decide what to read in-depth or to simply classify the texts. Topic
modelling is an automatic approach that attempts to extract the most important topics per text document.
The basic assumption of topic modelling \cite{Steyvers:ti} is that documents are created using a set of topics the authors want to describe
and discuss in the documents. The topics might, however, not be explicitly specified in the documents and remain only
implicitly in the heads of the authors. Nevertheless, for each topic, the authors still use certain words in the documents. Therefore, 
for this analysis, we say that a topic is formed by a set of related words. Hence,
there are probabilities with which certain words appear in the context of several topics. Topic modelling makes use of this by aiming
to extract these probabilities and thereby recreating the topics. Figure~\ref{fig:topic-modelling} shows the whole process
from document creation based on topics and the subsequent topic modelling to rediscover the topics. Hence, the user
of topic modelling does not have to specify any topics to look for in the documents, but they are extracted from the text.

Mathematically, we need an algorithm which is able to group the words extracted from documents into probable topics.
The most common one used is \emph{Latent Dirichlet Allocation} (LDA)~\cite{blei03} but there are others to choose from. The 
concrete algorithm is mostly uninteresting for the user of the topic modelling method, because all algorithms are not exact. 
An exact algorithm is impossible to define as the goal (the topics) are not clearly defined. The algorithms typically
start by assigning topics to words randomly and then use Bayesian probability to incrementally refine the assignment.
Finally, when an equilibrium is reached and the assignments cannot be improved, we have the most probable topics
for the corpus of documents.

The uses for topic modelling of software engineering data are vast. For example, if we have a larger body of existing
specifications to which our software has to conform, we can generate a network of documents based on the topics they share.
Fortunately, there is open and usable tool support for building topic maps. \emph{Mallet}\footnote{\url{http://mallet.cs.umass.edu/}}
is written in Java and allows users either to run it using the command-line or include it into their own analysis software
using an API. Hence, topic modelling is a useful tool for getting an overview of a large corpus of documents.

By applying topic modelling to our running example of RFC specifications using Mallet, we can reconstruct several useful
topics related to HTTP and IMAP. Figure~\ref{fig:topic-modelling} also illustrates some of the rediscovered topics. The top 
topic for the URI specification contains \emph{server}, \emph{message},
\emph{client} and \emph{standards}. For the HTTP specification, we get the terms \emph{request}, \emph{response}, \emph{header} and
\emph{http}, and for the IMAP specification \emph{command}, \emph{mailbox}, \emph{message}, and \emph{http}. Not every word
in each topic is helpful. Some can even be confusing, such as \emph{http} in IMAP, but most of the topics give a good idea what 
the specification is about. Furthermore, each document can have more than one topic. Figure~\ref{fig:topic-modelling} shows
this by the thickness of the arrows.

A simple alternative for small documents, which are not suitable for topic modelling, are word clouds (see also Section~\ref{sec:visualisation}) 
and counts. They cannot show semantic relationships between words but infer the importance of words by their frequencies. Available web tools, such as Voyant\footnote{\url{http://voyant-tools.org}},
can also show the context in which chosen words appear in the text and, thereby, provide an initial semantic flavour.

Topic modelling can only give a rough idea what are the main topics consisting of important words. The further
analysis and interpretation needs manual effort. Yet, especially for larger text corpora, topic modelling can be an
interesting pre-analysis before manual coding. The topics found can form initial ideas for coding, and we can mark
context in which they were found to be checked in detail by the coder.

\subsection{Part-of-Speech Tagging and Relationship Extraction}% (400 words, 2 fig)}
\label{sec:pos}

A way to further dig into the meaning of a large text corpus is to analyse its syntax in more
detail. A common first step for this is to annotate each word with its grammatical task. This is also
referred to as \emph{part-of-speech} (POS) tagging. In its simplest form,
it means extracting which word is a noun, verb or adjective. Contemporary POS taggers~\cite{brill00}
are able to annotate more, for example the tense of a verb. The taggers use machine learning
techniques to build models of languages to be able to do the annotations.

We see an example sentence from the HTTP 1.0 specification in Tab.~\ref{tab:pos}. We POS-tagged 
this sentence using the \emph{Stanford Log-linear Part-Of-Speech Tagger}~\cite{toutanova00}.
The tags are attached to each word with a ``\_'' as separator. They use a common abbreviation
system for the POS. For example, ``DT'' is a determiner and ``NN'' is a singular noun. The full
list of abbreviations can be found in \cite{marcus93}.

\begin{table}[htb]
\caption{A POS-tagged sentence from RFC 1945}
\centering
\begin{tabular}{p{14cm}}
\hline
\textbf{Original sentence from RFC 1945}\\
\hline
  The Hypertext Transfer Protocol (HTTP) is an application-level
   protocol with the lightness and speed necessary for distributed,
   collaborative, hypermedia information systems.\\
\hline
\textbf{POS-tagged sentence}\\
\hline
The\emph{\_DT} Hypertext\emph{\_NNP} Transfer\emph{\_NN} Protocol\emph{\_NNP -LRB-\_-LRB-} HTTP\emph{\_NNP -RRB-\_-RRB-} is\emph{\_VBZ} an\emph{\_DT} application-level\emph{\_JJ} protocol\emph{\_NN} with\emph{\_IN} the\emph{\_DT} lightness\emph{\_NN} and\emph{\_CC} speed\emph{\_NN} necessary\emph{\_JJ} for\emph{\_IN} distributed\emph{\_VBN} ,\_, collaborative\emph{\_JJ} ,\_, hypermedia\emph{\_NN} information\emph{\_NN} systems\emph{\_NNS} .\_. \\
\hline
\end{tabular}
\label{tab:pos}
\end{table}

This allows us to extract the main nouns which probably form a part of the domain concepts of the
specified system. In the example, if we combine consecutive nouns, we will find ``Hypertext Transfer Protocol'', 
``HTTP'', ``protocol'', ``lightness'', ``speed'' and ``hypermedia information systems''. These nouns 
capture already a lot of the main concepts of the specification. We can further qualify them with adjectives.
For example, the specification is not only about hypermedia information systems but about ``collaborative''
hypermedia information systems. Yet, we also see a problem in this kind of analysis in this example. The word ``distributed''
is tagged as verb (VBN) instead of an adjective which would probably distort an automated analysis.

\begin{figure}[!hbt]
\centering
  \includegraphics[width=0.7\textwidth]{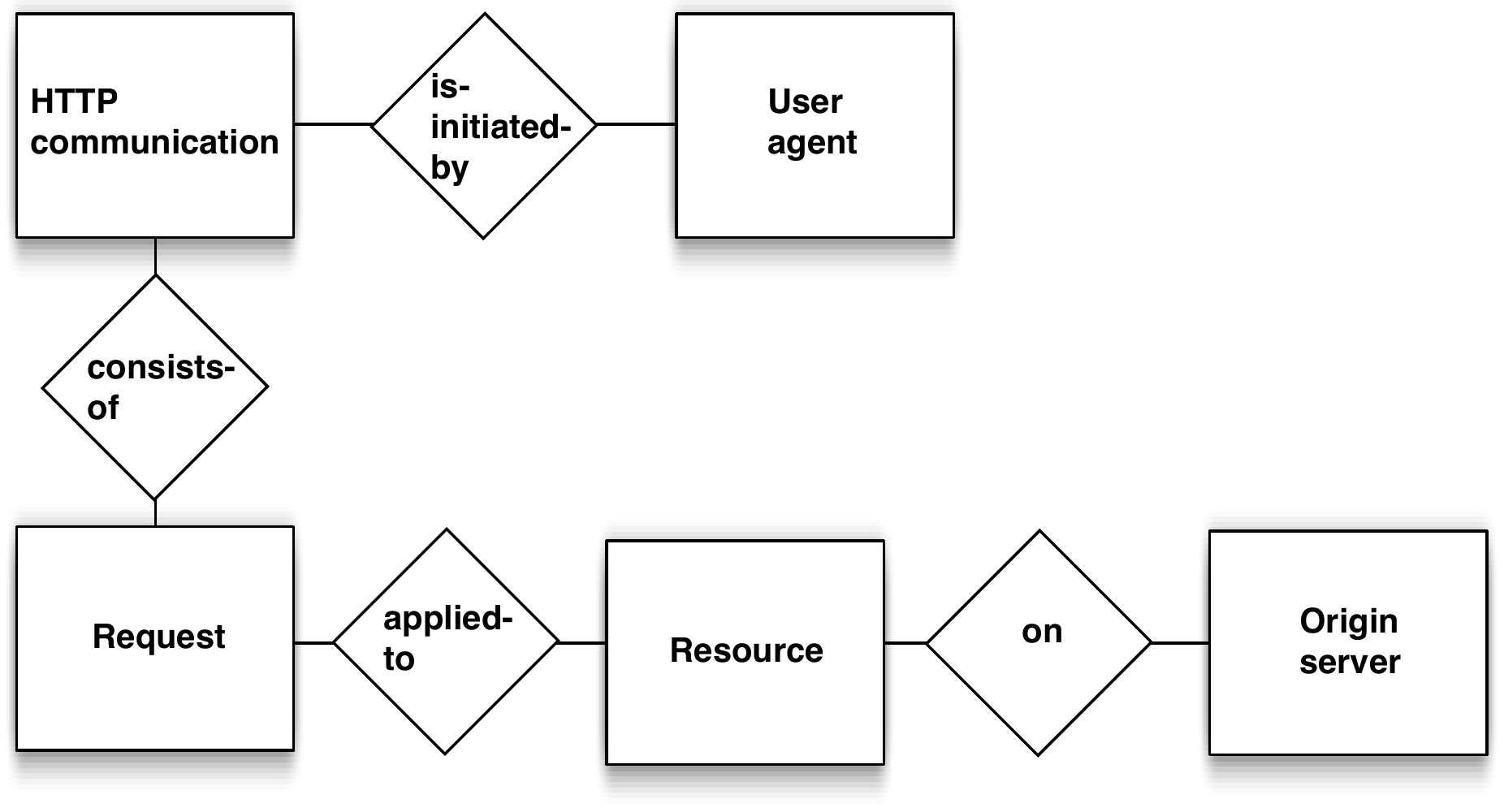}\\
  \caption{An ER diagram derived from a sentence from RFC 1945}\label{fig:er}
\end{figure}

The possibilities of further analysis having the POS tags in a text are very broad. A concrete example of
exploiting this in the context of software engineering is to extract domain information from a specification
to generate design models or code skeletons. For example, Chen investigated the structure
of English prose specifications to create entity-relationship (ER) diagrams~\cite{Chen:1983vk}. Put very simply, he maps nouns to entities and
verbs to relationships. For example, the sentence
\begin{quotation}
Most HTTP communication is initiated by a user agent and consists of
a request to be applied to a resource on some origin server.
\end{quotation}
from RFC 1945 can be transformed into the five entities \emph{HTTP communication}, \emph{user agent}, \emph{request},
\emph{resource} and \emph{origin server}. They are connected by the relationships \emph{is initiated by}, \emph{consists of},
\emph{to be applied to} and \emph{on}. The resulting ER diagram is shown in Figure~\ref{fig:er}.

Similar approaches were proposed by Abbott~\cite{Abbott:1983bf} to create a first skeleton of Ada code based on a
textual specification or by Kof~\cite{Kof05} who has built an ontology from a textual specification to infer initial
component diagrams and message sequence charts. All these approaches can help to bridge the gap from textual 
specifications to further, more formal artefacts. Yet, a multitude of other applications of POS tagging are also possible on other kinds
of textual software engineering artefacts. For example, we can assess the quality of requirements specifications
by detecting requirements smells such as passive voice \cite{femmer14}. Furthermore, it can help as a preprocessing
step in manual coding by highlighting nouns, adjectives, and verbs with different colours to quickly grasp the main
concepts.

\subsection{N-Grams}% (350 words, 1 fig)}

Computational linguists are looking for ways to predict what a next word in a sentence could be. One way to achieve this
is by looking at the immediately preceding words. ``On the basis of having looked at a lot of text, we know
which words tend to follow other words.''~\cite{manning99} Hence, we need a way of grouping these preceding words. A popular way
is to construct a model grouping words having the same preceding $n-1$ words. This model is then called an $n$-gram model.
An n-gram is a contiguous sequence of $n$ words in a piece of text. The $n$-gram-based analysis of texts is not aiming at abstracting 
the content of the texts but to categorise or predict attributes of them. 

$n$-gram models have received a lot of interest in the last years. Part of the interest comes
from the availability of the \emph{Google Ngram Viewer}\footnote{\url{http://books.google.com/ngrams}} for the books digitised by Google.
It can show the percentage of an n-gram in relation to all $n$-grams of all books per year. For example, for the 2-gram
``software engineering'', we can see a spike in the early 1990s and a mostly flat line since the 2000s. So, one
application of n-grams is to compare how frequently words occur together in different texts or over time.

Another interesting application of $n$-grams is for categorising texts into their languages. Based on already learned models for
different languages, it can indicate in which language is a given text. Imagine, your company policy is to write all documents
in English, including specifications and  code comments. Then, an analyser using the n-gram models could look for non-english 
text in code and other documents. Another example is to automatically classify specification chapters into technical content and
domain content. A useful tool in that context is the \emph{Java Text Categorizing Library} (JTCL)\footnote{\url{http://textcat.sourceforge.net}}.
It comes with $n$-gram models for a set of languages and is also capable of being trained for other categories. When we sent
the RFC specification documents we use as running example into JTCL, it correctly classified them as being written in English.

There are further uses of $n$-gram models in software engineering. Hindle et al.~\cite{Hindle:2012cz} investigated the naturalness of source 
code. They built $n$-gram models for source code and then predicted how to complete code snippets similar to auto completion in modern IDEs.
Allamanis and Sutton~\cite{Allamanis:2013cf} built on the work by Hindle et al.\ and created $n$-gram language models of the whole
Java corpus available in GitHub. Using these models, they derived a new kind of complexity metric based on how difficult it is to predict 
the sequence of the given code. The intuition behind this complexity metric is that complex source code is also hard to predict. Hence, the
worse the prediction matches the actual source code piece, the more complex it is. These applications are not ripe for wide-spread
industrial use but show the potential of the analysis technique.

\subsection{Clone Detection}
\label{sec:clone-detection}

Clone detection is a static analysis technique born in code analysis but usable on all kinds of
texts. It is a powerful technique to get an impression of the syntactic redundancy of a software, and it is highly automated at the same time.

\paragraph{What Is a Clone?}

A clone is a part of a software development artifact that appears
more than once. Most of the clone detection today concentrates on code
clones but cloning can happen in any artifact. In code, it is usually the result
of a normal practice during programming: Developers realize that they have implemented
something similar somewhere else. They copy that part of the code and adapt it
so that it fits their new requirements. So far, it is not problematic, because we
expect that the developer will perform a refactoring afterwards to remove the introduced
redundancy. Often, however, this does not happen, either because of time
pressure or because the developer is not even aware that this can be a 
problem.

A developer most often does not create an exact copy of the code piece but changes
some identifiers or even adds or removes some lines of code. The notion of a
clone incorporates that, too. To identify something as a clone, we allow 
normalization to some degree, such as different identifiers and reformatting.
If complete statements (or lines of text) have been changed, added or deleted, we speak of
\emph{gapped clones}. In clone detection, we then have to calibrate how large this gap should be allowed to be. If it is set too large, at some point everything will be a clone. Yet, it can be very interesting to see clones with 3 to 5
lines difference. 

As mentioned above, clone detection is not restricted to source code. If the particular
detection approach permits it, we can find clones in any kind of text. For example, we have applied our
clone detection tool on textual requirements specifications and found plenty of requirements
clones. We will discuss this study in detail in Section~\ref{sec:clone-study}. This works
because clone detection in the tool ConQAT\footnote{\url{http://www.conqat.org/}} is
implemented based on tokens which we can find in any text. Only normalization cannot
be done because we cannot differentiate identifiers.

\paragraph{Impact of Cloning}

It is still questioned today in research if cloning is really a 
problem~\cite{Kapser:2008en} while Martin states that ``Duplication may be the root of
all evil in software'' ~\cite{martin08}. There are many factors influencing the effects of cloning.
In our studies, however, we found two clearly negative impacts of cloning.

First, it is undeniable that the size of the software becomes larger than it
needs to be. Every copy of text adds to this increase in size which could
be avoided often by simple refactorings. There are border cases where a
rearranging would add so much additional complexity that the positive effect
of avoiding the clone would be compensated. In the vast majority of cases,
however, a refactoring would support the readability of the text. The size
of a software codebase is correlated to the effort needed to be spent to read, change,
review, and test it. The review effort increases
massively, and the reviewers become frustrated because they have to
read a lot of similar text.

Second, we found that cloning can also lead to unnecessary faults. We
conducted an empirical study~\cite{juergens09}
with several industrial as well as an open source system in which we
particularly investigated the gapped clones in the code of those systems. We reviewed
all found gapped clones and checked whether the differences were intentional
and whether they constitute a fault. We found that almost every other
unintentional inconsistency (gap) between clones was a fault. This way,
we identified 107 faults in five systems that have been in operation for
several years. Hence, cloning is also a serious threat to program correctness. 

\paragraph{Clone Detection Techniques}
There are various techniques and tools to detect clones in different artifacts~\cite{koschke2007survey}. 
They range from token-based comparison \cite{Juergens2009Clone} over the analysis of abstract syntax
trees \cite{2007_JiangL_decard} to more semantics-close analyses such as memory states~\cite{Kim:2011jh}.
In the following example, we work with the tool ConQAT mentioned above. 
It is applied in many practical environments to regularly check for cloning in code and other
artifacts. The measure we use to analyse cloning is predominantly \emph{clone coverage}
which describes the probability that a randomly chosen line of text exists more
than once (as a clone) in the system. In our studies, we often found code clone coverage
values for source code between 20~\% and 30~\% but also 70~\% to 80~\% is not rare. The best
code usually has single digit values in clone coverage. For other artifacts, it is more difficult
to give average values but we have also found clone coverage up to 50~\% in requirements
specifications~\cite{Juergens:2010iw}.

In general, false positives tend to be a big problem in static analysis. For clone detection,
however, we have been able to get rid of this problem almost completely. It requires
a small degree of calibration of the clone detection approach for a context but then
the remaining false positive rates are neglectable. ConQAT, for example, provides
black listing of single clones and can take regular expressions describing text to be ignored, such 
as copyright headers or generated code.
Finally, we use several of the visualizations the dashboard tool ConQAT provides
to control cloning: A trend chart shows if cloning is increasing or a tree map shows
us which parts of our systems are affected more or less strongly by cloning.

\paragraph{Running Example}
We now run the standard text clone detection of ConQAT on the HTTP and IMAP RFCs. By inspecting
the found clones, we find several false positives, for example the copyright header which is similar
in all documents. While these are copies, we do not care because we do not have to read them in detail
or compare them for differences. We ignore the headers by giving ConQAT a corresponding regular
expression. Figure~\ref{fig:clone-example} shows an example of a remaining clone in RFC 2616. It describes 
two variations of HTTP types which are mostly the same. A manual check of what is the same and what is 
different between the two clone instances would be boring and error-prone. Hence, clone analysis could be 
used to point to text parts that should be removed and, thereby, make
understanding of the specifications easier.

\begin{figure}[htb]
%org.apache.hivemind.Reverser.reverse(String)
%org.apache.commons.lang.StringUtils.reverse(String)
\tiny
\begin{tabular}{@{}p{\dimexpr0.5\linewidth-\tabcolsep}% can be up to 0.5\linewidth 
                 | p{\dimexpr0.43\linewidth-\tabcolsep}@{}}
\begin{lstlisting}
Media Type name:         message
Media subtype name:      http
Required parameters:     none
Optional parameters:     version, msgtype
    version: The HTTP-Version number of the enclosed message
               (e.g., "1.1"). If not present, the version can be
               determined from the first line of the body.
    msgtype: The message type -- "request" or "response". If not
               present, the type can be determined from the first
               line of the body.
Encoding considerations:
\end{lstlisting}
&
\begin{lstlisting}
Media Type name:         application
Media subtype name:      http
Required parameters:     none
Optional parameters:     version, msgtype
    version: The HTTP-Version number of the enclosed messages
                 (e.g., "1.1"). If not present, the version can be
                 determined from the first line of the body.
    msgtype: The message type -- "request" or "response". If not
                 present, the type can be determined from the first
                 line of the body.
Encoding considerations:
\end{lstlisting}
\end{tabular}
\caption{A text clone of RFC 2616 \label{fig:clone-example}}
\end{figure}

\begin{figure}[!hbt]
\centering
  \includegraphics[width=0.7\textwidth]{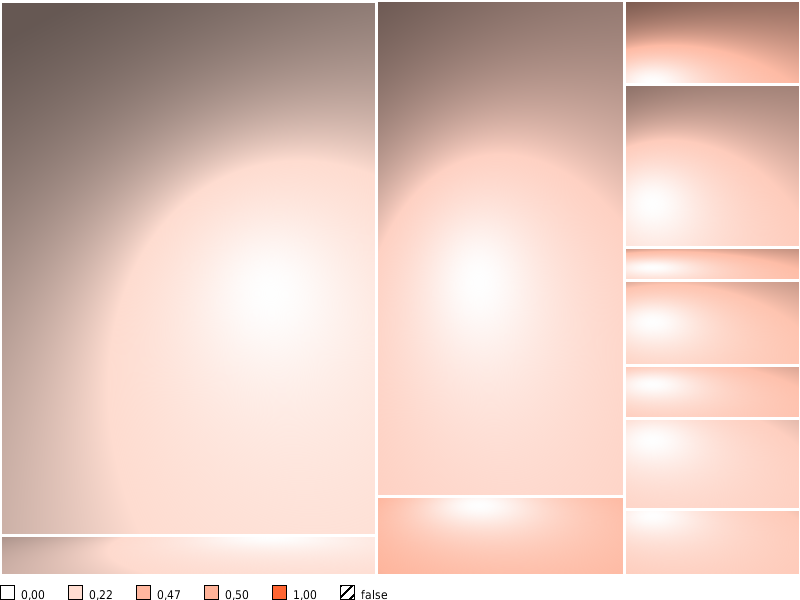}\\
  \caption{A tree map of the cloning in the analysed RFCs}\label{fig:treemap}
\end{figure}

Overall, we have a clone coverage of 27.7~\% over all 11 analysed documents from the HTTP and
IMAP specifications. The clone coverage
is rather evenly distributed as shown in the tree map in Figure~\ref{fig:treemap}. The tree map displays
each file as a rectangle. The size of the rectangle is relative to the size of the file and the color indicates
the clone coverage. The more red a rectangle is, the higher the clone coverage for this file. This can give
us a fast overview of even a large number of documents.

Manual coding as well as automatic techniques, such as POS tagging or topic modelling, aim at 
providing a concise abstraction of the text. We can use it to describe, summarise, and better
understand the text. Clone detection, however, has the goal to describe the redundancy created
by copy\&paste in the text. Hence, a possibility is to use clone detection as a first analysis
step to exclude clones from the further analysis. Simply copied texts will otherwise distort the
other analyses.

\subsection{Visualization}% (300 words, 4 figs)}
\label{sec:visualisation}

``A picture is worth a thousand words.'' is a clich\'e but a graphical visualization can help strongly in understanding
the results of text analyses. In the previous sections, we have already seen a tree map in Figure~\ref{fig:treemap}
to get a quick overview of cloning in different text files. It could be used to visualize the distribution of all kinds 
of metrics over files. A further visualization was the ER diagram in Figure~\ref{fig:er} which shows the domain concepts
in a text in an abstract form well known by computer scientists.

The support for analysts by visualization is a very active research area, often denoted by the term \emph{visual analytics}.
Also for visualising unstructured textual data, several new methods and tools have appeared over the last years. 
Alencar et al.~\cite{Alencar:2012it} give a good overview of this area. We will base our discussion on their work and only highlight
three exemplary visualizations.

\begin{figure}[!hbt]
\centering
  \includegraphics[width=0.6\textwidth]{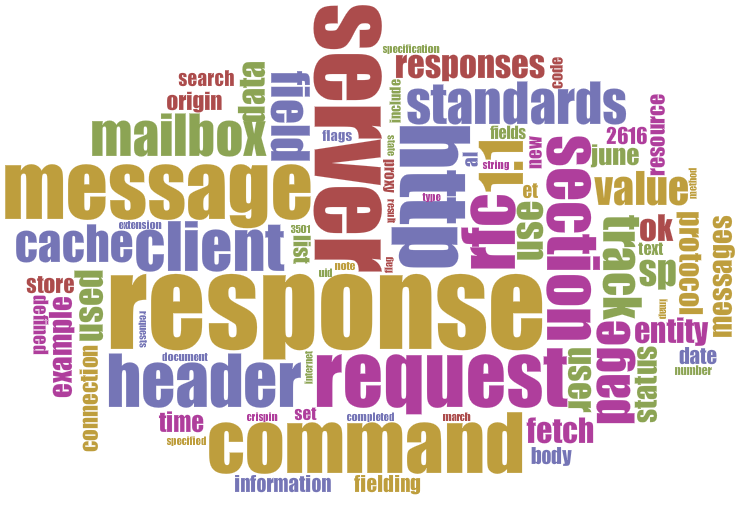}\\
  \caption{A word cloud of HTTP and IMAP}\label{fig:word-cloud}
\end{figure}

\paragraph{Word Cloud} 
A nowadays well known, simple but effective visualization of text is called \emph{word cloud}, \emph{tag crowd}\footnote{\url{http://tagcrowd.com/}}
or \emph{wordle}\footnote{\url{http://www.wordle.net/}}. Different implementations give different concrete visualizations but the
idea is always to extract the most frequent words and show them together with the size of each word in relation to its frequency. Our example
of the HTTP and IMAP RFCs gives us the word cloud in Figure~\ref{fig:word-cloud}.
A useful tool for that is Voyant\footnote{\url{http://voyant-tools.org}} which we have already mentioned in the context
of topic modelling (Section~\ref{sec:topic-modelling}). It can create a word cloud out of a text corpus and allows interactively
to change its appearance, include stop word lists (to avoid having ``the'' as the largest word) and click onto any word to see
its frequency and context. The word cloud in Figure~\ref{fig:word-cloud} was created using Voyant. A word cloud can be a
good first step for finding the most important terms in a set of texts. For example, it could be an input into a manual coding
process for identifying a set of prior codes likely to appear often in the texts.

\begin{figure}[!hbt]
\centering
  \includegraphics[width=0.9\textwidth]{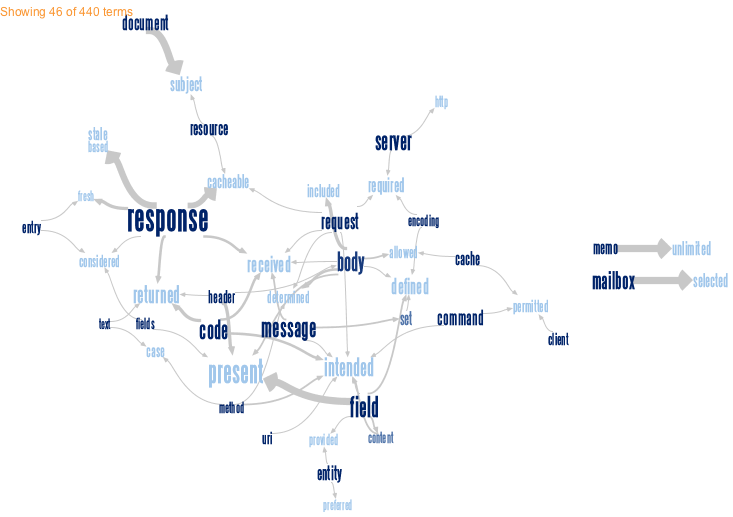}\\
  \caption{A Phrase Net of HTTP and IMAP with the relationship ``is'' created with IBM's Many Eyes}\label{fig:phrase-net}
\end{figure}

\paragraph{Phrase Net}
A comparable visualization that adds a bit more semantics is a \emph{Phrase Net}~\cite{vanham09}. It not only presents very
frequent words but also their relationship to other frequent words. The kind of relationship can be configured: for example, the connection
of both words by another word such as ``the'' or ``is'' can be a relationship. Figure~\ref{fig:phrase-net} shows a Phrase Net 
for the RFC corpus with the selected
relationship ``is''. It shows frequent words, the larger the more frequent, as well as arrows between words that are connected by an ``is''. The
arrow becomes thicker the more frequent the relationship occurs in the text. For example, the RFCs contain often the word ``response'' -- as we already
saw in the word cloud -- but we additionally see that ``response'' is frequently connected by ``is'' to ``state-based'' and ``cacheable''.
If the data can be publicly accessible, the IVM Many Eyes\footnote{\url{http://www-958.ibm.com/software/analytics/labs/manyeyes/}} system is 
an easy tool to create Phrase Nets. It is a more complicated visualization than word clouds but also gives more information.
It could be used as an alternative to create domain models as with part-of-speech tagging (Section~\ref{sec:pos}) or to check
and extend those domain models. Furthermore, it could provide a more comprehensive input to manual coding as it not
only contains single words but important relationships between words.

\paragraph{Temporal Change} 
A next step in visualising textual data has been to introduce a further dimension in the visualization such as the change 
over time in texts. This can be interesting for the analysis of several versions
of a document or survey results collected regularly. We can see changes in the interest in different topics, for example. 
Havre et al.~\cite{havre02} propose \emph{ThemeRiver} to visualise topics as streams horizontally over time with the thickness of the stream
in relation to the strength of the topic at that point in time. Chi et al.~\cite{cui11} extend this with their approach \emph{TextFlow} which adds 
specific events extracted from the texts such as the birth or death of a topic to the streams. At present, there is no tool available to perform
this kind of analysis. 

\begin{figure}[!hbt]
\centering
  \includegraphics[width=0.8\textwidth]{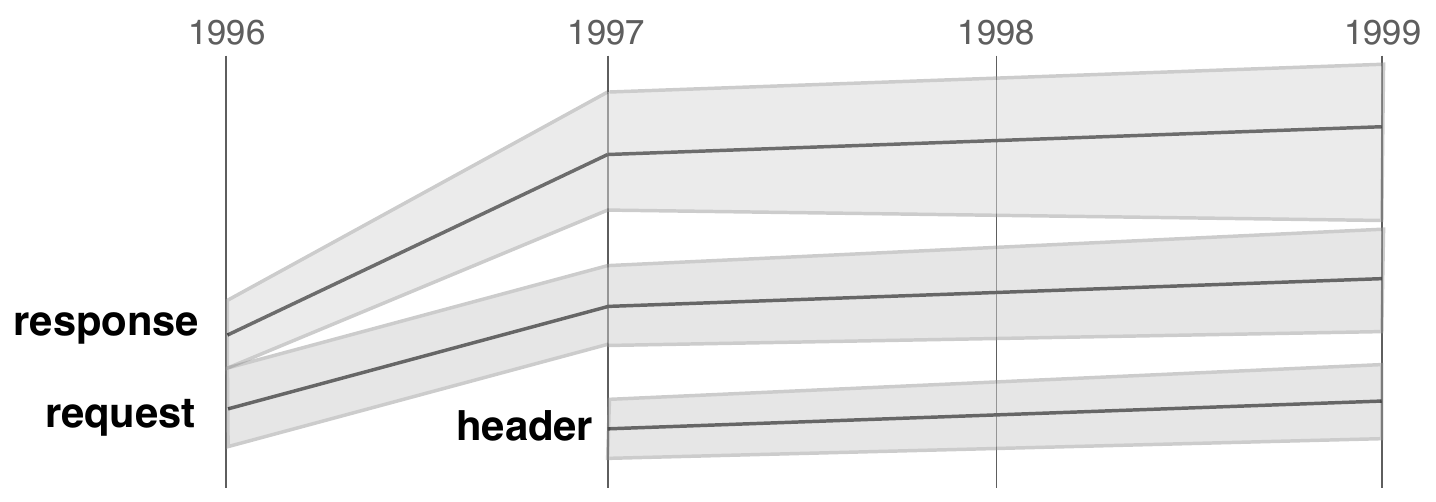}\\
  \caption{A TextFlow sketch of the different HTTP RFCs over time}\label{fig:textflow}
\end{figure}

We sketched a text flow of the different versions of the RFC on HTTP in Figure~\ref{fig:textflow} to show the concept. We see
that in 1996, the stream for ``request'' is bigger than for ``response'' and, hence, was more often used. This changes in the later versions. The
word ``header'' comes only in as a very frequent word in 1997 and continues to increase in importance in 1999.

\section{Two Industrial Studies}
\label{sec:IndustrialStudies}

We further illustrate the application of text analytics in software engineering by two industrial studies: first, a survey on requirements
engineering we manually coded and analysed, and, second, clone detection on requirements engineering which we combined
with manual coding of the found requirements clones.

\subsection{NaPiRE: A Requirements Engineering Survey}% (DMF) (1000 words, 2 tables, 2 fig)}
\label{sec:NaPiRE}

We conducted this survey study in 2013 as a collaboration between TU M\"unchen and the University of Stuttgart. We have been working with
industrial partners on requirements engineering (RE) for several years and had a subjective understanding of typical problems in this area.
Yet, we often stumbled on the fact that there is no more general and systematic investigation of the state of the practice and contemporary
problems of performing requirements engineering in practice. Therefore, we developed a study design and questionnaire to tackle this challenge
called \emph{Naming the Pain in Requirements Engineering} (NaPiRE).
While you are not likely to perform the same study, the way we analysed the free-text answers to our open questions is applicable to
any kind of survey. You can find more information on the complete survey in \cite{MW2013a,MW13b} and on the website: \url{http://www.re-survey.org/}.

\paragraph{Goals and Design}
Our long-term research objective is to establish an open and generalisable set of empirical findings about practical problems and needs in RE that allows 
us to steer future research in a problem-driven manner. To this end, we want to conduct a continuously and independently replicated, globally distributed 
survey on RE that investigates the state of the practice and trends including industrial expectations, status quo, experienced problems and what effects those 
problems have. The survey presented in the following describes the first run of our survey in Germany.

Based on these goals, we developed a set of research questions and derived a study design and questionnaire. For most of the aspects we were
interested in, we designed closed questions that can be analysed with common quantitative analyses from descriptive statistics. Often
we used the Likert-scale from ``I fully agree'' to ``I do not agree at all'' to let the survey respondents rate their experiences. We complemented 
closed questions often with an open question to let the respondents voice additional opinions. At the end of the questionnaire,
we asked open questions about the personal experiences with requirements engineering in their projects. Our design included a manual analysis
using manual coding (Section~\ref{sec:manual-coding}) of all the textual data we would get from the open questions. We wanted to derive an improved
and extended understanding of potential problems which we then include in the closed questions of the next survey run.

\paragraph{Example Questions and Answers}
Let us look at two examples of open questions we asked in the questionnaire and some answers we got. The first question we discuss is:
\begin{quotation}
If you use an internal improvement standard and not an external one, what where the
reasons?
\end{quotation}
The context was that we first asked about normative standards, defined by external parties, that they use for improving their requirements
engineering practices. An example of such a standard would be the Capability Maturity Model Integration (CMMI) of the U.S.\ Software Engineering
Institute and adaptations of that standard for requirements engineering. We were interested in how satisfied the respondents
are with such standards and, in turn, why they did not use them. Answers included ``We want to live our own agility.'', ``We do not use any standard''
and ``I am not convinced of the external standards.''

The second question we look at provoked longer answers from most respondents. It is also the one where we could get the most from with our
manual coding later. It reads:
\begin{quotation}
Considering your personally experienced most critical problems (selected in the previous question), how do these problems manifest themselves in the process, e.g., in requests
for changes?
\end{quotation}
We had presented a set of problems we encountered in practice before in the previous question. Now, we wanted to better understand what problems the respondents consider most critical. From this, we wanted to learn about the context of the problems, potential
causes and candidates for new problems. The answers we got included ``Requirements emerge and change priority once the system is deployed. Again, this is no problem 
but the environment. It becomes a problem if you're not prepared to deal with it.'' and ``Hard negotiations regarding the CR/Bug question, mostly leading 
to bad relationship with the customer on project lead level.''

\paragraph{Coding Process and Challenges}
For analysing the free-text answers, we followed the manual coding procedure as introduced in Section~\ref{sec:manual-coding}. However, we already 
have a predefined set of codes (given RE problems) for which we want to know how the participants see their implications. For this reason,  we have to 
deviate our procedure from the standard procedure and rely on a mix of bottom-up and top-down. We start with selective coding and build the core 
category with two sub-categories, namely \emph{RE problems} with a set of codes each representing one RE problem and \emph{Implications} which then groups 
the codes defined for the answers given by the participants. For the second category, we conduct open coding and axial coding for the answers until reaching a 
saturation for a hierarchy of (sub-)categories, codes and relationships. 

\begin{figure}[!htb]
\centering
  \includegraphics[width=1\textwidth]{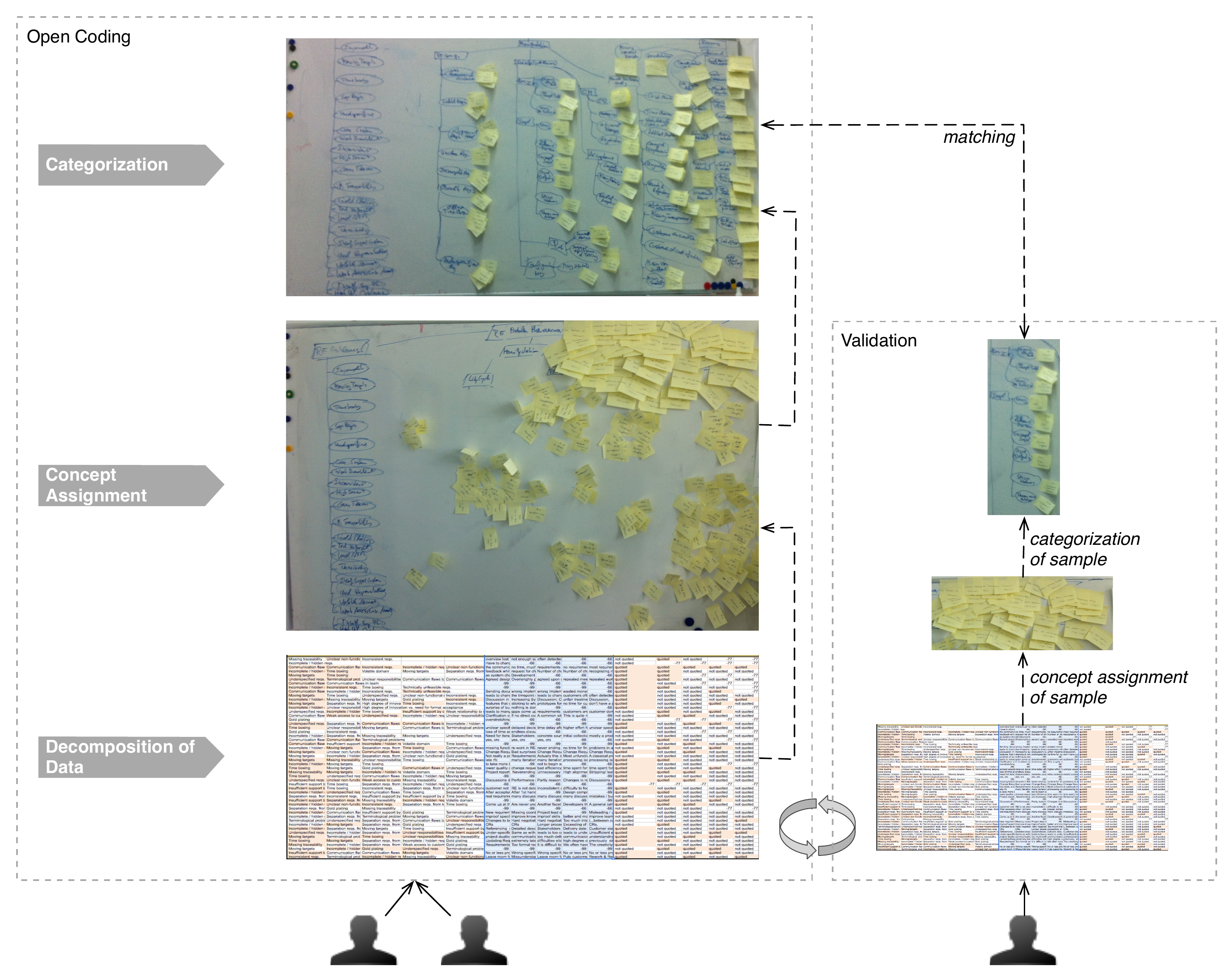}\\
  \caption{Open coding and validation procedure}
  \label{fig.OpenCodingNaPiRE}
\end{figure}

During the coding process, we had to tackle several challenges. One was the lack of appropriate tool support for manual coding, especially when working in distributed environments. Another one was the missing possibility to validate the results by getting feedback from the respondents. Figure~\ref{fig.OpenCodingNaPiRE} sketches our procedure we followed during manual coding.

For this reason, we relied on analyst triangulation during the open coding step as this was essentially the step which most depended on subjectivity (during interpretation of the answers to the open questions). During this open coding step, we decomposed the data relying on spread sheets and worked with paper cards where we also denoted the rationale for selected codes. In a third step, we arranged the cards according to categories by using a whiteboard. A third analyst then repeated, as a validation step, independently the open coding process on a sample.

\paragraph{Coding Results}
Due to the resulting complexity in the given answers and resulting coding scheme, we describe the results step-wise.
To this end, we first introduce the full results from the open coding, followed by the full results of the axial coding. In a last step, we present a condensed 
overall view of the results as a graph with a minimal saturation. We show only those results having a minimal occurrence in the answers to include only those in our theory.

\begin{figure}[!htb]
\centering
  \includegraphics[width=0.82\textwidth]{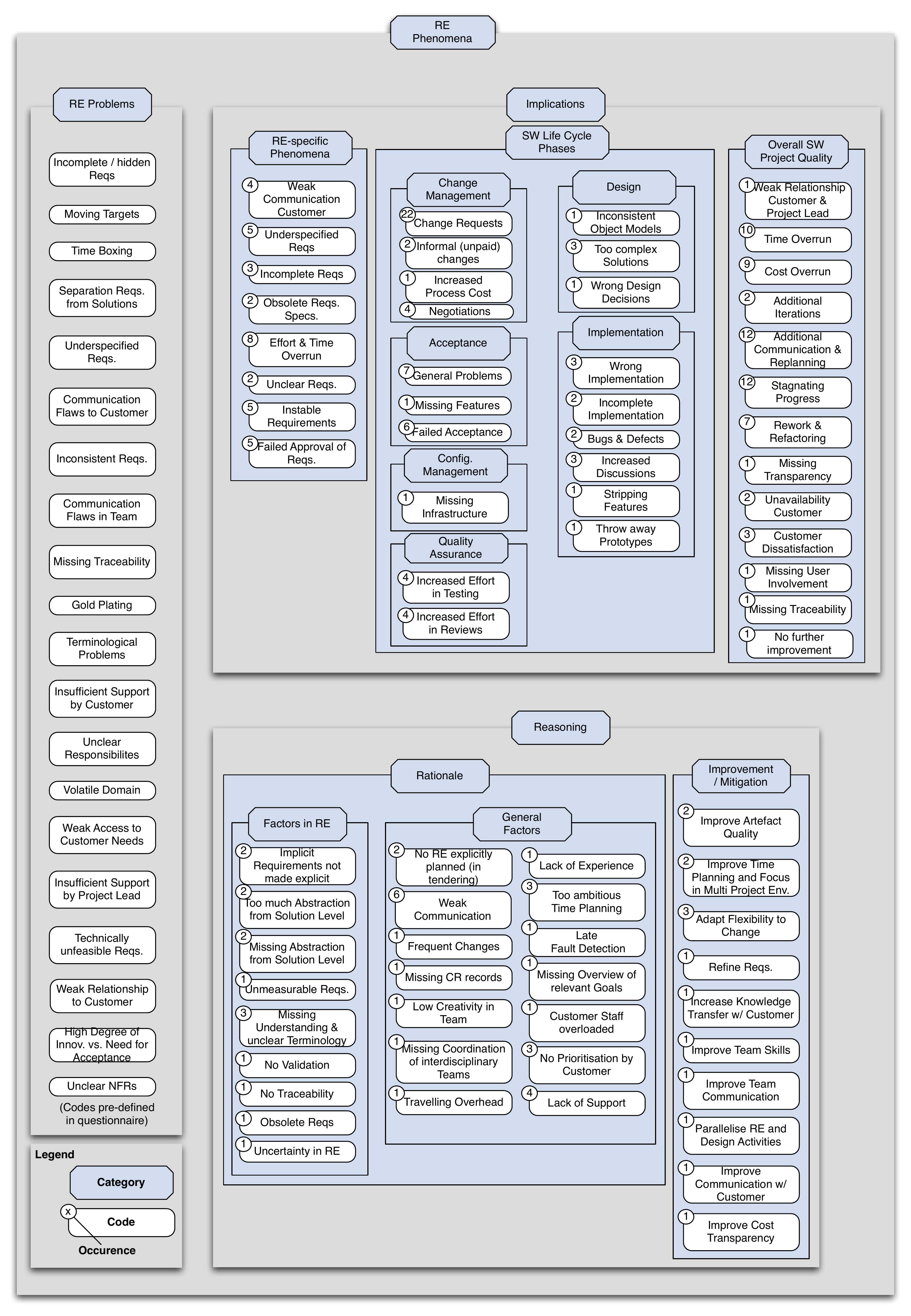}\\
  \caption{Categories and codes resulting from open coding}
  \label{fig.OpenCoding}
\end{figure}

%\subsubsection*{Full Results from Open Coding}
Figure~\ref{fig.OpenCoding} summarises the full results from the open coding. We distinguish a hierarchy of categories as an abstraction of those codes defined for
 the answers given in the questionnaire. For each code, we furthermore denote the number of occurrences. Not included in the coding are statements that cannot 
 be unambiguously allocated to a code, for example, the statement ``never ending story'' as an implication of the problem ``Incomplete / Hidden requirements''.

Given that we asked what implications the problems have, we would expect two top-level categories. The participants also stated, however, reasons for the 
problems and occasionally also how they would expect to mitigate the problem. As shown in Fig.~\ref{fig.OpenCoding}, we thus categorise the results into the 
pre-defined category \emph{RE Problems}, \emph{Implications}, and the additional category \emph{Reasoning}.

Regarding the implications, we distinguish three sub-categories. Consequences of problems in the RE phase itself, consequences to be seen in further phases of 
the SW life cycle other than RE and more abstract consequences on the overall project quality. The highest occurrence of statements is given for the code 
\emph{Change Request} being stated 22 times. Other codes resulted from only one statement, but they were unique, specific formulations that could not be merged 
with other statements. For instance, the code \emph{Weak Relationship Customer \& Project Lead} in the category \emph{Overall SW Project Quality} resulted from 
a statement which we could not allocate to another code without interpretation and potentially misinterpreting the statement (given that no validation with the respondents is possible).

Regarding the reasoning for given RE problems, we distinguish the category \emph{Rationale} as a justification of why particular problems occurred, and 
\emph{Improvement/Mitigation} for statements that suggested how to mitigate particular problems.  The first category can be further divided into \emph{Factors in RE} 
and \emph{General Factors}. 

Also here, we encountered very mixed statements including detailed ones we had to allocate to codes having in the end only one occurrence and vague statements we 
could accumulate with (consequently vague) codes. Subsequent original statements shall give an impression about the answers given:
\begin{compactitem}
\item[Code \emph{Missing Abstraction from Solution Level:}] ``Stakeholders like to discuss on solution level, not on requirements level. Developers think in solutions. The problem is: even Product Managers and Consultants do it.''
\item[Code \emph{No RE explicitly planned (in tendering):}] ``A common situation is to take part in a tender process -- where requirements are specified very abstract -- most of these tender processes do not include a refinement stage, as a supplier we are bound to fulfill vague requests from the initial documents.''
\item[Code \emph{Weak Communication:}] ``The communication to customer is done not by technicians, but by lawyers.''
\item[Code \emph{Too Ambitious Time Planning:}] ``Delivery date is known before requirements are clear.''
\item[Code \emph{Implicit Requirements not made explicit:}] ``Referencing  common sense  as a requirement basis.''
\item[Code \emph{Failed Acceptance:}] ``After acceptance testing failed, the hidden requirements came up and had to be fixed on an emergency level.''
\item[Code \emph{Missing Coordination of Interdisciplinary Teams:}] ``Missing coordination between different disciplines (electrical engineering, mechanical engineering, software etc.).''
\end{compactitem} 

%\subsubsection*{Full Results from Axial Coding}

The axial coding defines the relationships between the codes. As a consequence of the categories introduced in the previous section, we distinguish two types of relationships:
\begin{compactenum}
\item The consequences of given RE problems to the category \emph{Implications}, and
\item The consequences of the codes in the category \emph{Reasoning} to the RE problems including rationales and improvement suggestions.
\end{compactenum}

We cannot show the full results of the axial coding here. We further refrain from interpreting any transitive relationships from reasonings to implications because of the multiple input-/output-relationships between the codes of the different categories; for instance, while ``Too ambitious Time Planning'' was stated as an exclusive reason for ``Time Boxing'' as an RE problem, the problem ``Incomplete / hidden Requirements'' has multiple reasons as well as multiple consequences.

%\subsubsection*{Condensed Overall View with Minimal Saturation}

\begin{figure}[!htb]
\centering
  \includegraphics[width=1\textwidth]{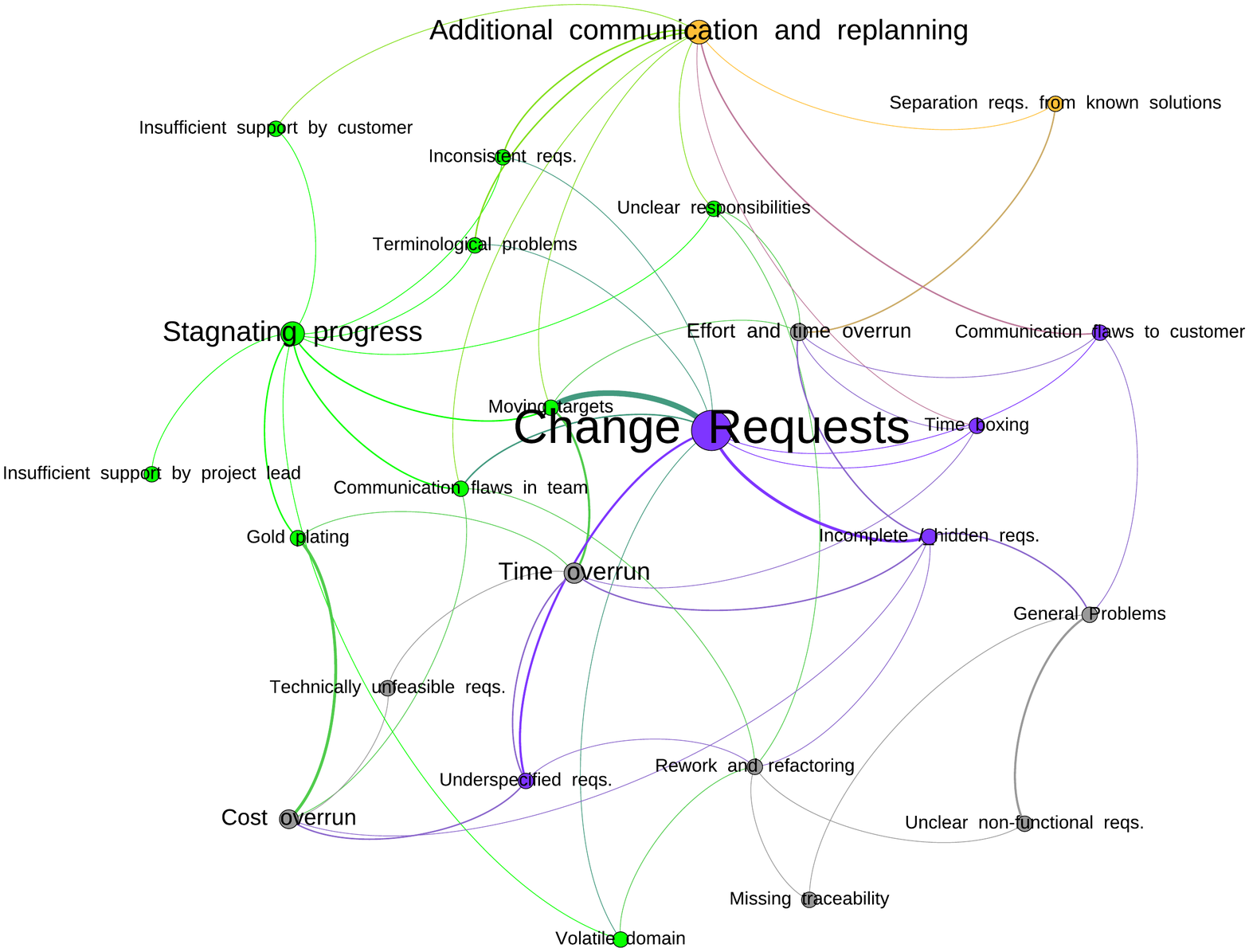}\\
  \caption{Condensed view on axial coding with minimal weighting of 7 in the nodes.}
  \label{fig.axialCodingCondensed}
\end{figure}

We are especially interested in a condensed result set that omits the codes and the dependencies with limited occurrences in corresponding statements. The reason is that we
need a result set with a minimal saturation to propose its integration into the questionnaire for the next run of the survey.  After testing the results with different values for the minimal 
occurrences, we defined a graph including only codes with a minimal occurrence level of 7. This resulting graph is shown in Fig.~\ref{fig.axialCodingCondensed}. Nodes represent 
a selection of codes and edges represent a selection of relationships between nodes.

With the chosen minimal occurrence level, the final graph does not include statements coded in the category \emph{Reasoning} leaving us with a selection of RE problems interconnected with their implications. The three nodes with the highest occurrence in their underlying statements are differently colored. Change requests (in the centre of the figure) are stated as the most frequent consequence of various RE problems such as time boxing or incomplete / hidden requirements. Additional communication and replanning (upper part of the figure) was another frequently stated consequence of interconnected RE problems, similar as a stagnating process (left side of the figure).

\subsection{Clone Detection in Requirements Specifications}% (SW) (1000 words, 1 table, 2 figs)}
\label{sec:clone-study}

We performed this case study in 2009 in a collaboration between TU M\"unchen and itestra GmbH as an exploratory study
on the extent of cloning in commercial requirements specifications. We had a subjective feeling that there is a lot of textual
redundancy in those specifications from the experiences with project partners but we had not investigated systematically before.
All details of the study can be found in \cite{Juergens:2010iw}.
It is an example for clone detection on natural language texts (Section~\ref{sec:clone-detection}) as well for manual 
coding (Section~\ref{sec:manual-coding}).

\paragraph{What Is the Problem with Clones in Specifications?}
Requirements specifications are a central artifact in most software development processes. They capture the goals of the
software to be developed and constitute the connection between the customer/user and the developers. Many call the specifications
the determining part for project success or failure. Yet, as with any other development artifact, requirements specifications contain
redundancy. Semantic redundancies are the source for many potential problems in a development project but are also extremely hard
to detect. Yet, there are also syntactic redundancies: Specifications are written by people who tend to copy (and adapt) text if they 
need similar things at different parts of the document or different documents. These syntactic redundancies can be found by clone detection
(see Section~\ref{sec:clone-detection}).

In our collaborations with various industry partners on their requirements engineering processes, we often found syntactic redundancies
in their specifications. They lead to various problems. The most direct is the sheer increased size of the specifications which, in turn,
leads to higher efforts for reading, reviewing and changing them. In addition, similar to code cloning \cite{juergens09}, redundancy can
introduce inconsistencies. The copies of the text drift apart over time as some copies are adapted, for example to changes in the
requirements of the customer, while others are forgotten. We now have conflicting requirements and the developers can introduce
faults into the software. Finally, an undesired effect of clones in requirements specifications is also that the developers introduce
redundancy by cloning the implementation or, even worse, develop the same functionality more than once.

\begin{figure}[hbt]
\centering
  \includegraphics[width=0.9\textwidth]{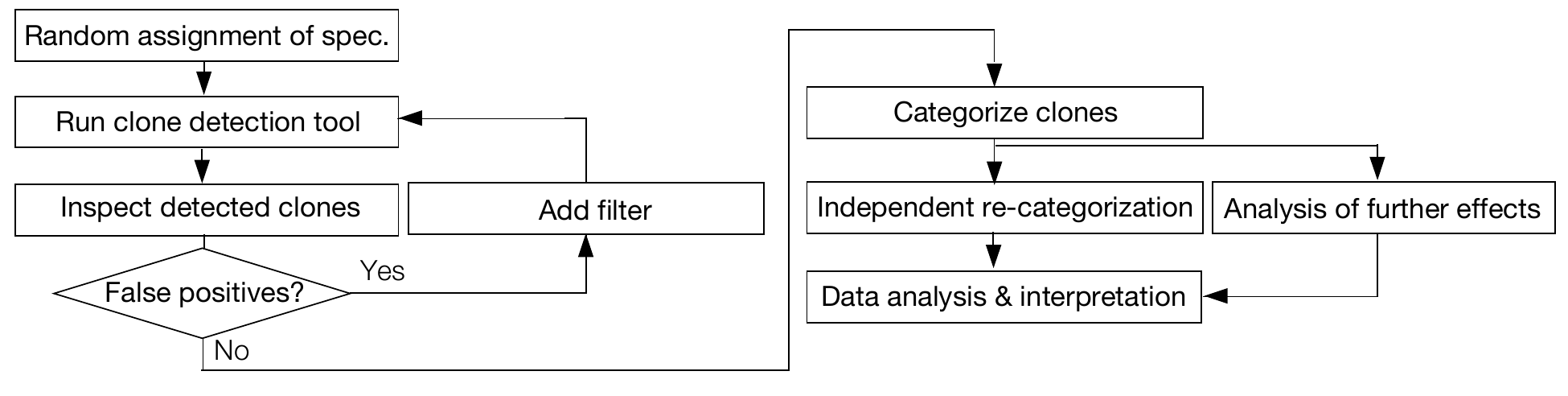}\\
  \caption{Our approach to analyse cloning in requirements specifications}\label{fig:cloning-design}
\end{figure}

\paragraph{Analysis Approach}
To better understand the actual extent of cloning in industrial requirements specifications, we designed a study in which
we use automated clone detection on a set of industrial specifications and then classify the found clones with manual coding. The former gives us
a quantification of the phenomenon and the latter a qualitative insight into what information is cloned. We worked in several 
research pairs and followed the process shown in Fig.~\ref{fig:cloning-design}.

We assembled a set of 28 requirements specifications of various length and for a wide variety of systems and domains.
We first assigned the specifications randomly among pairs of analysts. Each pair ran a clone detection without any filters
using the tool ConQAT. We then inspected the found clones for false positives (pieces of text reported as clones but not
actual redundancies). As expected, we found a few false positives such as copyright headers and footers. We added
corresponding filters in the form of regular expressions so that the false positives will be ignored. Afterwards, we ran the clone detection again and inspected
the results. This continued until we could not find any false positives in a random sample of the clones. This gave us the
quantitative results of the study.

Afterwards, we manually coded a random sample of the clones to form categories of the type of information that was cloned.
We had no pre-defined codes but developed them while inspecting the clones. As we did not need any larger theory, we
skipped axial coding (Section~\ref{sec:manual-coding}). The coding gave us a complete set of categories.
To validate the more subjectively developed categories, we made an independent re-coding of a sample of the
categorised clones and found a substantial agreement between the raters. Besides that, we also noted additional effects
such as the impact on the implementation. This gave us the qualitative results of the study.

\paragraph{Results of Automatic Analysis}
All the quantitative results of our study are shown in Tab.~\ref{tab:cloning-results}. The outcome is clear: There are many
industrial requirements specifications that contain cloning, several with high clone coverage values between 30~\% and 70~\%.
The third column gives the number of clone groups per specification. A clone group is a set of clones, the individual copies.
There are several specifications with more than 100 clone groups. Hence, there has been a lot of copy\&paste in these documents.
There are also several specifications, however, with no or almost no cloning. Therefore, it seems to be possible to create specifications
without copy\&paste.

\begin{table}[htb]
\caption{Automatic analysis results of cloning in requirements specifications\label{tab:cloning-results}}
\begin{center}
%{\small\sffamily
\begin{tabular}{lrrr}
\hline

& \multicolumn{1}{c}{\textbf{Clone}} & \multicolumn{1}{c}{\textbf{Clone}} &  \\
\textbf{Spec}  & \multicolumn{1}{c}{\textbf{cov.}} & \multicolumn{1}{c}{\textbf{groups}} & \multicolumn{1}{c}{\textbf{Clones}} \\
\hline
H & 71.6~\% & 71 & 360 \\
F & 51.1~\% & 50 & 162 \\
A & 35.0~\% & 259 & 914 \\
G & 22.1~\% & 60 & 262 \\
Y & 21.9~\% & 181 & 553 \\
L & 20.5~\% & 303 & 794 \\
Z & 19.6~\% & 50 & 117 \\
C & 18.5~\% & 37 & 88 \\
K & 18.1~\% & 19 & 55 \\
U & 15.5~\% & 85 & 237 \\
X & 12.4~\% & 21 & 45 \\
AB & 12.1~\% & 635 & 1818\\
V & 11.2~\% & 201 & 485 \\
B & 8.9~\% & 265 & 639 \\
N & 8.2~\% & 159 & 373 \\
D & 8.1~\% & 105 & 479 \\
P & 5.8~\% & 5 & 10 \\
I & 5.5~\% & 7 & 15 \\
AC & 5.4~\% & 65 & 148 \\
W & 2.0~\% & 14 & 31 \\
O & 1.9~\% & 8 & 16 \\
S & 1.6~\% & 11 & 27 \\
M & 1.2~\% & 11 & 23 \\
J & 1.0~\% & 1 & 2 \\
E & 0.9~\% & 6 & 12 \\
R & 0.7~\% & 2 & 4 \\
Q & 0.0~\% & 0 & 0 \\
T & 0.0~\% & 0 & 0 \\
\hline
Avg & 13.6~\% &  &   \\
$\Sigma$ &  & 2,631 & 7,669 \\\hline

\end{tabular}%}
\end{center}
\end{table}

 \paragraph{Results of Manual Classification}

\begin{figure}[h]
\centering
\includegraphics[width=.8\linewidth]{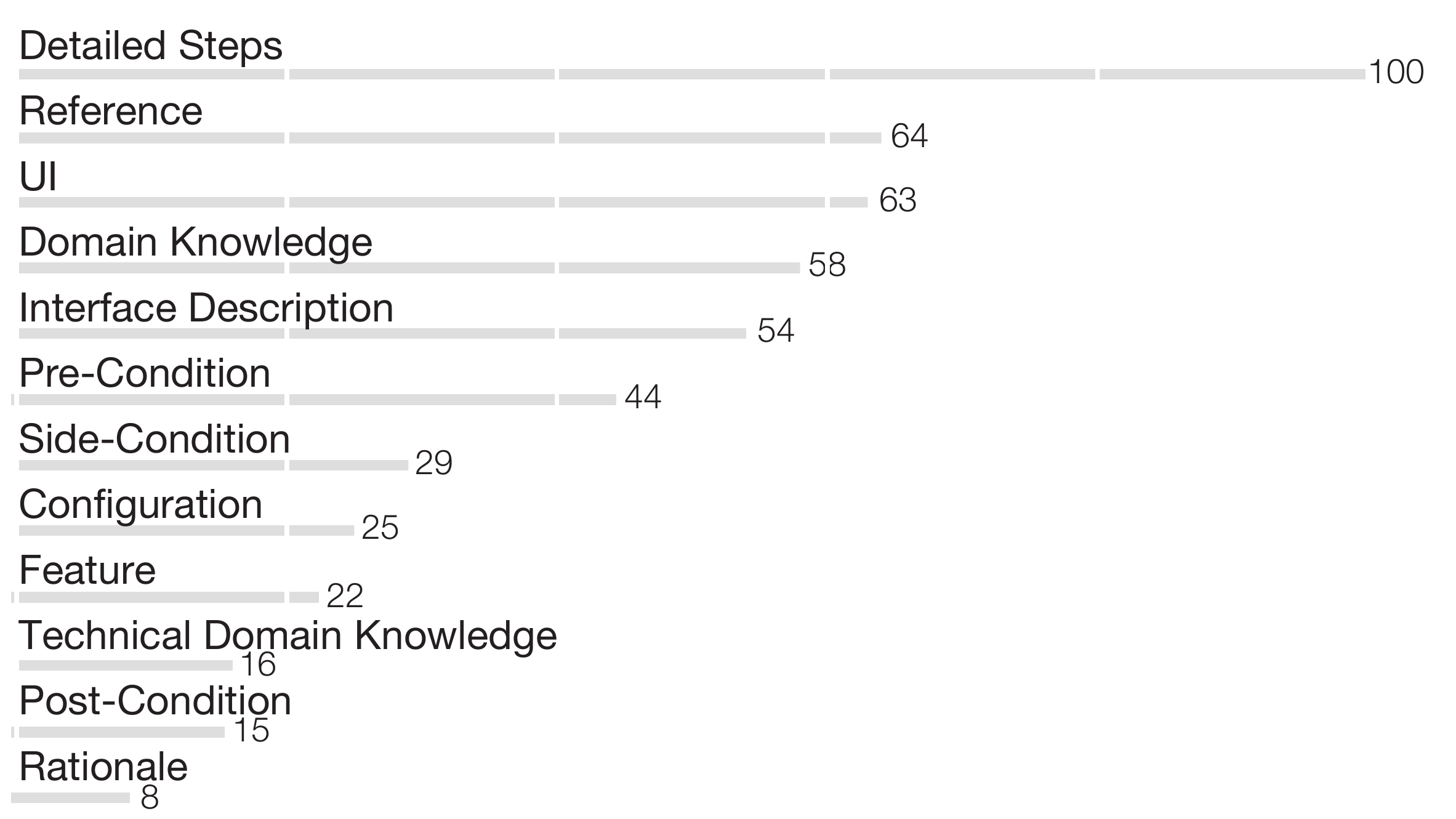}
\caption{Cloning: Found categories of requirements clones}
\label{fig:cloning-categorisation}
\end{figure}

The manual coding of a sample of clones resulted in 12 categories of cloned information
encountered. The categories we identified are described in Table~\ref{tab:cloning-categories}~\cite{Juergens:2010iw}.
Overall, we coded a sample of over 400 clone groups almost 500 times because we
sometimes assigned a clone group to more than one category, especially if the clones were
longer and, hence, contained different aspects. To better understand these different categories
and how they occur in practice, we quantified the results by counting the number of clone groups
per category in our sample (Fig.~\ref{fig:cloning-categorisation}). The highest
number of assigned codes are to the category ``Detailed Use Case Steps'' with 100
assignments. ``Reference'' (64) and ``UI'' (63) follow. The least number of
assignments are to the category ``Rationale'' (8).

\begin{table}[htp]
\caption{Descriptions of the categories of cloned information in requirements specifications}
\begin{center}
\begin{tabular}{rp{10cm}}
\hline
\textbf{Detailed Use Case Steps} & Description of one or more steps in a use case that
         specifies in detail how a user interacts with the system, such as the
         steps required to create a new customer account in a system.\\
\textbf{Reference} & Fragment in a requirements specification that
         refers to another document or another part of the same document. Examples
         are references in a use case to other use cases or to the corresponding
         business process.\\
\textbf{UI} & Information that refers to the
         (graphical) user interface. The specification of which buttons are visible on
         which screen is an example for this category.\\
\textbf{Domain Knowledge} & Information about the application
         domain of the software. An example are details about what is part of an
         insurance contract for a software that manages insurance contracts.\\
\textbf{Interface Description} & Data and message definitions that describe the interface of a component, function, or system.
         An example is the definition of messages on a bus system that a component
         reads and writes.\\
\textbf{Pre-Condition} & A condition that has to hold before something else can happen.
         A common example are pre-conditions for the execution of a specific use case.\\
\textbf{Side-Condition} & Condition that describes the status that has to hold during the
         execution of something. An example is that a user has to remain logged
         in during the execution of a certain functionality.\\
\textbf{Configuration} & Explicit settings for configuring
         the described component or system. An
         example are timing parameters for configuring a transmission protocol.\\
\textbf{Feature} & Description of a piece of functionality of the system on a high level
         of abstraction.\\
\textbf{Technical Domain Knowledge} & Information
         about the used technology for the solution and the technical
         environment of the system, for example used bus systems in an embedded system.\\
\textbf{Post-Condition} & Condition that describes what has to hold after something
         has been finished. Analogous to the pre-conditions, post-conditions are usually
         part of use cases to describe the system state after the use case execution.\\
\textbf{Rationale} & Justification of a requirement. An example is the explicit demand
         by a certain user group.\\
\hline
\end{tabular}
\end{center}
\label{tab:cloning-categories}
\end{table}%

Overall, this study was a beneficial combination of manual and automatic text analyses to better
understand the extent and type of cloning in requirements specifications. The automatic analysis
has the advantage that we could integrate it into a regular quality analysis, for example contained
in the nightly build of the corresponding system. This way, introduced clones could be detected and
removed early and easily. Nevertheless, it is interesting to regularly also inspect the clones and
categorise them to understand if there appear new types of clones.

\section{Summary}% (SW) (300 words)}

Textual data constitutes the majority of data that is generated in a software project. Yet, we often
do not make use of the information and insights contained in this textual data. The problem is that
many analysis methods are focused on quantitative data. There are various possibilities of manual
as well as automatic analyses now available that help us in employing textual data in better understanding
our software projects.

In this chapter, we saw first manual coding to analyse any kind of textual
data. We assign different types of codes to the text to abstract and interpret it. This is a highly subjective
task which needs appropriate means such as triangulation of analysts to make it more objective. Yet, it
can be flexibly applied and allows the analysts to bring in their own expertise in the data analysis. The
biggest problem, however, is the large amount of effort necessary for the analysis.

Therefore, we discussed a sample of automatic analyses available, mostly with easily accessible tool
support. For example, clone detection is an easy and viable means to detect any syntactic redundancy
in textual software artifacts. Topic modelling is another example that can help us to investigate dependencies
between documents or quickly get an overview of topics contained in the documents. Visualizations can
greatly support all of these automatic analyses to make them easier to comprehend, especially for large
text corpora.

The research on text analytics is still very active and we expect to see many more innovations
that we will be able to exploit also for analysing textual project data from software projects. The
possibilities are huge.

\bibliographystyle{model1-num-names}
\bibliography{literature}

\end{document}